\documentclass[]{interact}

\usepackage{amsfonts}
\usepackage{graphicx}
\usepackage{amsopn}
\usepackage[numbers,sort&compress]{natbib}

\theoremstyle{plain}

\theoremstyle{definition}

\theoremstyle{remark}

\usepackage{amsmath}
\usepackage{amssymb}
\usepackage{enumerate}
\usepackage{url} 
\usepackage{algorithm}
\usepackage{algpseudocode}
\usepackage[caption=false]{subfig}
\usepackage{tikz}
\usepackage{hyperref}

\hypersetup{breaklinks=true,
            bookmarks=true,
            pdfauthor={J. Derek Tucker () and John R. Lewis () and Caleb King () and Sebastian Kurtek (Sandia National Laboratories and The Ohio State University)},
             pdfkeywords = {Compositional noise, functional data analysis, functional tolerance
bounds, functional Principal Component Analysis},
            pdftitle={A Geometric Approach for Computing Tolerance Bounds for Elastic
Functional Data},
            colorlinks=true,
            citecolor=blue,
            urlcolor=blue,
            linkcolor=magenta,
            pdfborder={0 0 0}}

\providecommand{\tightlist}{%
  \setlength{\itemsep}{0pt}\setlength{\parskip}{0pt}}

  \def\x{{\mathbf{x}}}

  \def\F{{\mathcal{F}}}
  
  \def\R{{\mathcal{R}}}

  \newcommand{\s}{\ensuremath{\mathbb{S}}}
  \newcommand{\real}{\mathbb{R}}
  
  \newcommand{\T}{\ensuremath{\mathsf{T}}}
  
  \newcommand{\ltwo}{\ensuremath{\mathbb{L}^2}}
  \newcommand{\inner}[2]{\left\langle#1,#2 \right\rangle}
  \numberwithin{equation}{section}

  \usetikzlibrary{arrows,shapes,positioning,calc,fadings,decorations.pathreplacing,backgrounds}
  \tikzset{%
    >=latex, 
    inner sep=0pt,%
    outer sep=2pt,%
    mark coordinate/.style={inner sep=0pt,outer sep=0pt, draw=blue!90,fill=blue!90,minimum size=4pt,circle},
    mark coordinate2/.style={inner sep=0pt,outer sep=0pt, draw=red!90,fill=red!90,minimum size=4pt,circle}%
  }

  \newcommand\pgfmathsinandcos[3]{%
    \pgfmathsetmacro{#1}{sin (#3)}%
    \pgfmathsetmacro{#2}{cos (#3)}%
  }
  \newcommand\LongitudePlane[3][current plane]{%
    \pgfmathsinandcos\sinEl\cosEl{#2} 
    \pgfmathsinandcos\sint\cost{#3} 
    \tikzset{#1/.style={cm={\cost,\sint*\sinEl,0,\cosEl, (0,0)}}}
  }
  \newcommand\LatitudePlane[3][current plane]{%
    \pgfmathsinandcos\sinEl\cosEl{#2} 
    \pgfmathsinandcos\sint\cost{#3} 
    \pgfmathsetmacro{\yshift}{\cosEl*\sint}
    \tikzset{#1/.style={cm={\cost,0,0,\cost*\sinEl, (0,\yshift)}}} %
  }
  \newcommand\DrawLongitudeCircle[2][1]{
    \LongitudePlane{\angEl}{#2}
    \tikzset{current plane/.prefix style={scale=#1}}
    \pgfmathsetmacro{\angVis}{atan (sin (#2)*cos (\angEl)/sin (\angEl))} %
    \draw[current plane,black!70] (\angVis:1) arc (\angVis:\angVis+180:1);
  }
  \newcommand\DrawLatitudeCircle[2][1]{
    \LatitudePlane{\angEl}{#2}
    \tikzset{current plane/.prefix style={scale=#1}}
    \pgfmathsetmacro{\sinVis}{sin (#2)/cos (#2)*sin (\angEl)/cos (\angEl)}
    \pgfmathsetmacro{\angVis}{asin (min (1,max (\sinVis,-1)))}
    \draw[current plane,black!80] (\angVis:1) arc (\angVis:-\angVis-180:1);
  }

\begin{document}

  \title{A Geometric Approach for Computing Tolerance Bounds for Elastic Functional Data}

  \author{
  \name{J. Derek Tucker\textsuperscript{a}\thanks{CONTACT J. Derek Tucker. Email: jdtuck@sandia.gov}, John R Lewis\textsuperscript{a}, Caleb King\textsuperscript{a}, and Sebastian Kurtek\textsuperscript{b}}
  \affil{\textsuperscript{a}Statistical Sciences, Sandia National Laboratories, Albuquerque, NM} 
  \affil{\textsuperscript{b}Department of Statistics, The Ohio State University, Columbus, OH}
  }

  \maketitle

  \begin{abstract}
    We develop a method for constructing tolerance bounds for functional data with random warping variability. In particular, we define a generative, probabilistic model for the amplitude and phase components of such observations, which parsimoniously characterizes variability in the baseline data. Based on the proposed model, we define two different types of tolerance bounds that are able to measure both types of variability, and as a result, identify when the data has gone beyond the bounds of amplitude and/or phase. The first functional tolerance bounds are computed via a bootstrap procedure on the geometric space of amplitude and phase functions. The second functional tolerance bounds utilize functional Principal Component Analysis to construct a tolerance factor. This work is motivated by two main applications: process control and disease monitoring. The problem of statistical analysis and modeling of functional data in process control is important in determining when a production has moved beyond a baseline. Similarly, in biomedical applications, doctors use long, approximately periodic signals (such as the electrocardiogram) to diagnose and monitor diseases. In this context, it is desirable to identify abnormalities in these signals. We additionally consider a simulated example to assess our approach and compare it to two existing methods.
  \end{abstract}

  \begin{keywords}
    Compositional noise; functional data analysis; functional tolerance
    bounds; functional Principal Component Analysis
  \end{keywords}

\noindent  \hypertarget{introduction}{%
\section{Introduction}\label{introduction}}

A significant amount of data collected in biomedical applications,
process monitoring and reliability engineering is in the form of
functions where each data object is a collection of data points over
some index (e.g., time or frequency). In these applications, the goal is
often to provide estimates on the range in which a certain proportion of
the population of functions is expected to fall while accounting for
sampling uncertainty. With this goal in mind, this work focuses on
developing theoretically sound methods for constructing tolerance bounds
for functional data. Tolerance bounds are confidence bounds on quantiles
and can be used to construct ranges within which `there is
\((1-\alpha)100\%\) confidence that \((1-p)100\%\) of the population of
functional data falls' \cite{Hahn:2011}. For an equal-tailed
tolerance bound, the upper bound is an upper confidence bound on the
\(1-p/2\)-quantile, while the lower bound is a lower confidence bound on
the \(p/2\)-quantile. Such a notion of tolerance bounds is important in
many different applications.

In applied settings that involve collection and analysis of functional data,
it is common to ignore the dataset's functional nature and only extract
key scalar features on which to base inferences. For example, in
monitoring of an electrocardiogram of a patient, it is often desirable
to identify periods when a heart beat is outside of normal rhythm. Key
scalar features in this case could be the periodicity of the
electrocardiogram or the time between heartbeats. The statistical
analysis would then proceed by making inferential statements, such as
constructing tolerance bounds, on the key features only. Common examples
of other mathematical features extracted for such analyses include local
maxima and minima, the number of peaks, or a rate of change at a
particular point on the function. As an alternative to extracting and
analyzing a finite set of key features, the functional data can be first
discretized, and then treated as a finite vector. In this approach, many
tools from standard multivariate analysis, including Principal Component Analysis (PCA), become available. However, standard multivariate data analysis approaches ignore the intrinsic infinite-dimensional nature of the data, as well as the strong dependence between neighboring points on the functions. On the other hand, we aim to develop theory/methods that respect these important properties of functional data, and discretize only at the end of the overall process as a necessary step for computer implementation. Another important aspect of the data ignored by standard techniques that discretize at the outset of the analysis is potential horizontal (warping) variation, explained in the following paragraph. Ignoring such variation can make the analysis of functional data less meaningful.

\begin{figure}[!t]

{\centering \includegraphics{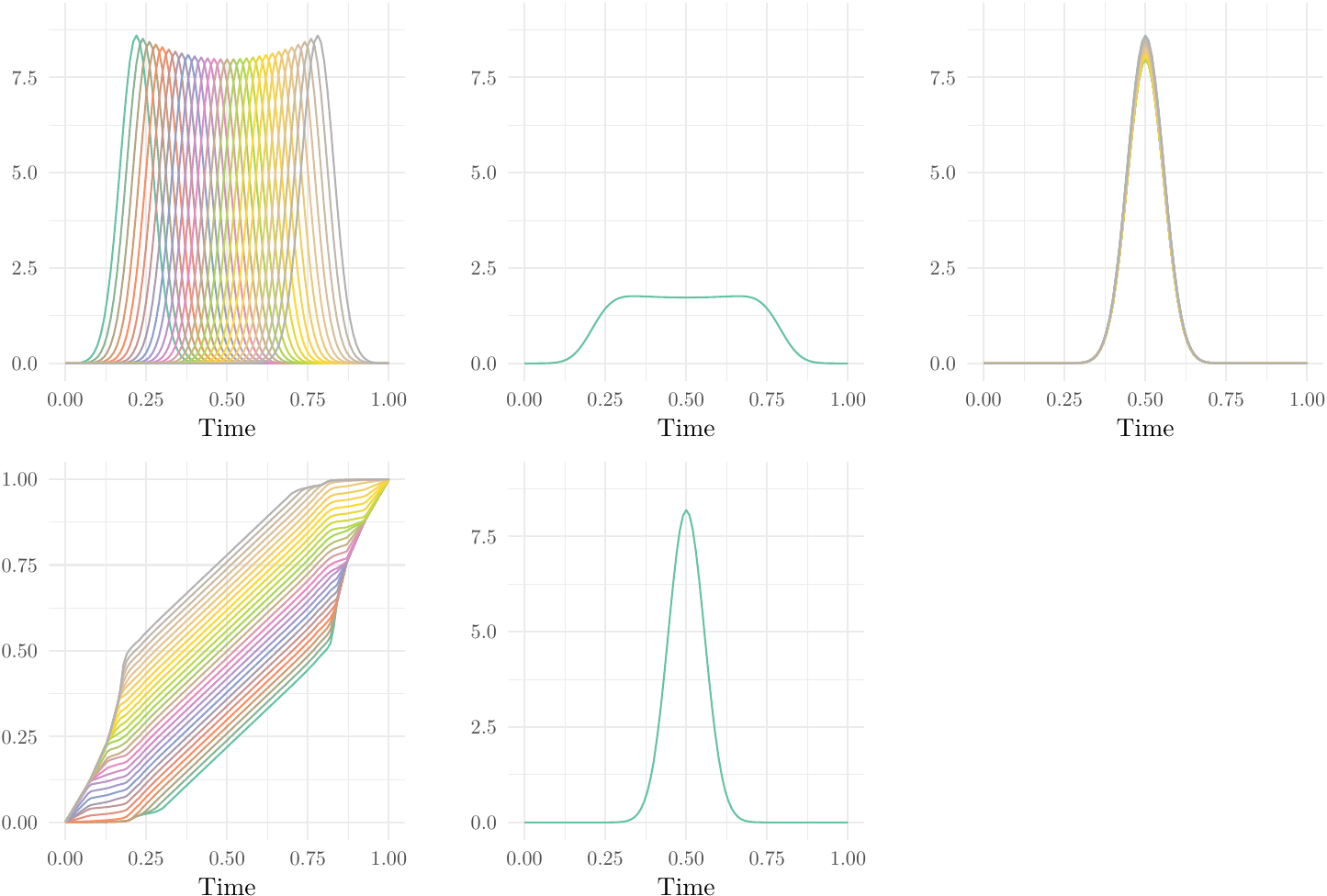}

}

\caption{Demonstration of amplitude and phase variability in functional data. Top left: Original functions. Top middle: Cross-sectional mean without alignment. Top right: Aligned functions (amplitude). Bottom left: Warping functions (phase). Bottom middle: Cross-sectional mean after alignment.}\label{fig:toy_example}
\end{figure}

Alternatively, there has been considerable effort in statistics to
develop methods that can analyze functional data objects without significant loss of
information due to summarization. Such methodology is known as functional data analysis and
has a rich history in Statistics. An excellent introduction to this
field is given in several books including \cite{ramsay-silverman-2005},
\cite{Ferraty:2006:NFD:1203444} and \cite{FSDA}. An interesting
aspect of most functional data is that the underlying variability can be
ascribed to two sources. These two sources are termed the amplitude (or
\(y\) or vertical) variability and the phase (or \(x\) or horizontal or
warping) variability. Capturing these two sources of variability is
crucial when modeling and monitoring functional data in a process
control architecture, and can greatly affect the construction of
tolerance bounds. In this work, we refer to functional data that
contains both amplitude and phase variability as \emph{elastic}. This
important concept is illustrated in Figure \ref{fig:toy_example} through
a simulated example. The observed functions are generated according to
the equation \(y_i(t) = z_i e^{-(t-a_i)^2/2}\),
\(t \in [0, 1], ~i=1,2,\dots, 29\), where \(z_{i}\) are \emph{i.i.d.}
\(\mathcal{N}(1, (0.05)^2)\) and \(a_i\) are \emph{i.i.d.}
\(\mathcal{N}(0, (1.25)^2)\). The top left panel shows the simulated
functions; each sample function is unimodal with slight variability in
the height of the peak and large variability in its placement. The
relative heights of the peak can be attributed to the amplitude
variability, while the different locations of the peak constitute the
phase variability. The cross-sectional (pointwise) mean of this data is
shown in the top middle panel. This mean ignores the phase variability
which results in averaging out of the main unimodal amplitude feature. If tolerance bounds were to be constructed using this
cross-sectional approach, the intrinsic shape of the data would be lost, and the
bounds would not capture the true underlying variability in the data.
Alternatively, the phase variability can be accounted for by first
aligning the functions. As an example, the top right panel shows
time-aligned functions. The alignment involves a transformation of the
horizontal axis via warping functions shown in the bottom left panel.
The aligned functions capture the amplitude variability while the
warping functions capture the phase variability. The cross-sectional
mean of the aligned functions (amplitude) is shown in the last panel,
where the sharp unimodal structure of the original data is retained. The
tolerance bounds developed in this work provide bounds that maintain the
shape of the data by accounting for both directions of variability.

\hypertarget{past-work-and-contributions}{%
\subsection{Past Work and
Contributions}\label{past-work-and-contributions}}

Recently, Storlie et al. \cite{storlie-2013} developed a method to test the
shape of a population of curves using a B-Spline basis, and a
hierarchical Gaussian process approach to form confidence intervals. Rathnayake and Choudhary \cite{Rathnayake:2016} developed tolerance bounds for functional data using functional Principal Component Analysis (fPCA).
Further, Sun and Genton \cite{Sun:2011} developed a boxplot display for functional data,
which provides a nice visualization technique for a sample of functions.
This approach can also detect functional outliers. A drawback of all of the aforementioned methods is that they do not account for potential  phase variability in functional data, i.e., they assume that the data either (1) does not need to be
aligned, or (2) has already been aligned in a pre-processing step (generally using some criterion that is unrelated to subsequent data analysis).
The first assumption is unrealistic in process control and biomedical applications, while the second approach results in suboptimal solutions due to the disjointedness of the alignment and data analysis procedures. A more systematic approach is to
develop methods that build the alignment step into the statistical procedure of interest.

A couple of papers in recent literature have taken into account the phase variability in the application of
monitoring functional data. Lewis et al. \cite{Lewis:2017} expanded upon the
generative model presented in \cite{tucker-wu-srivastava:2013}, and used a
bootstrap approach to generate tolerance bounds. However, the phase
variability in this approach was discarded, and as a result, the generated tolerance bounds
did not maintain the shape of the original data. Grasso et al.
\cite{grasso:2016} developed an idea similar to Storlie et al. \cite{storlie-2013}
for process monitoring. In this work, they built the alignment of the
functional data into their procedure, where they used a parametric model for warping functions in conjunction with fPCA.
However, the \(\ltwo\) metric they used for alignment and construction
of the subsequent bounds has serious theoretical limitations (e.g., the
pinching effect) as described in \cite{marron2015}. Additionally, the
use of a parametric model for warping functions may not be flexible
enough to achieve good alignment in general applications.

In a recent paper, Xie et al. \cite{XieKurtek:2016} developed an alternative
visualization approach to the method of \cite{Sun:2011}, that directly
takes warping variability into account in the construction of the
boxplot displays. Their method is based on the Riemannian geometry of
the amplitude and phase representation spaces and builds upon the
general elastic functional data analysis framework presented in
\cite{srivastava-etal-JASA:2011},
\cite{kurtek-wu-srivastava-NIPS:2011}, and
\cite{tucker-wu-srivastava:2013}. Our approach also builds upon that
work by generalizing to arbitrary quantile estimates while accounting
for sampling uncertainty.

In this paper, we present two methods for computing tolerance bounds for
functional data observed under random warping variability. First, using
a modification of the joint fPCA approach developed in \cite{Lee:2017},
we create a generative model for amplitude-phase functions. One can then
sample from this model and use the bootstrap approach to construct
tolerance bounds on the amplitude and phase components separately. The
benefit of this approach is that the tolerance bounds maintain the shape
of the original data while capturing both amplitude and phase
variability. This approach is similar to \cite{Rathnayake:2016}, but their method fails to account for the phase
variability, which greatly affects the structure of the tolerance
bounds.

The second approach also uses the previously mentioned joint fPCA
method, and constructs the tolerance bounds in the joint amplitude-phase
coefficient space (after projecting onto the lower dimensional space
spanned by the eigenvectors of the covariance). To do this, a
multivariate Gaussian model is assumed for joint amplitude-phase fPCA
coefficients and a tolerance factor is computed. A new function, which
one would like to test, can be projected onto the same coefficient space
and its tolerance score can be compared to the tolerance factor. This
approach is similar in spirit to the elastic functional statistical
process control (fSPC) presented in \cite{tucker:2016}.

The rest of this paper is organized as follows. In Section \ref{pca},
we review the relevant material from elastic functional data analysis
and develop a joint amplitude-phase fPCA model. Section
\ref{functional-tolerance-bounds} describes the two methods for
constructing tolerance bounds for elastic functional data. In Sections
\ref{simulations} and \ref{application-to-real-data}, we report the
results of applying the proposed approach to a simulated dataset and two
real datasets from different application domains. Finally, we close with
a brief summary and some ideas for future work in Section
\ref{conclusion}.

\hypertarget{pca}{%
\section{Combined Phase-Amplitude fPCA}\label{pca}}

We begin by giving a short review of the combined phase-amplitude fPCA
method of \cite{Lee:2017}, with a slight modification which will be
described clearly in later sections. Their method is based on the
functional data analysis approach outlined in
\cite{srivastava-etal-JASA:2011},
\cite{kurtek-wu-srivastava-NIPS:2011} and
\cite{tucker-wu-srivastava:2013}; see those references for more details.

Let \(f\) be a real-valued function with the domain \([0,1]\); this
domain can be easily generalized to any other compact subinterval of
\(\real\). For concreteness, only functions that are absolutely
continuous on \([0,1]\) will be considered and we let \(\F\) denote the
set of all such functions. In practice, since the observed data are
discrete anyway, this assumption is not a restriction. Also, let
\(\Gamma\) be the set of orientation-preserving diffeomorphisms of the
unit interval \([0,1]\):
\(\Gamma = \{\gamma: [0,1] \rightarrow [0,1] |~\gamma(0) = 0,~\gamma(1)=1,\gamma~\textnormal{is a diffeomorphism} \}\).
Elements of \(\Gamma\) play the role of warping functions. For any
\(f \in \F\) and \(\gamma \in \Gamma\), the composition
\(f \circ \gamma\) denotes the time warping of \(f\) by \(\gamma\). With
the composition operation, the set \(\Gamma\) is a Lie group with the
identity element \(\gamma_{id}(t) = t\). This is an important
observation since the group structure of \(\Gamma\) is seldom utilized
in past papers on functional data analysis.

As described in \cite{tucker-wu-srivastava:2013}, there are two metrics
to measure the amplitude and phase variability of functions. These
metrics are proper distances, one on the quotient space \(\F/\Gamma\)
(i.e., amplitude) and the other on the group \(\Gamma\) (i.e., phase).
The amplitude or \(y\)-distance for any two functions
\(f_1,\ f_2 \in \F\) is defined to be \begin{equation}
d_a(f_1, f_2) = \inf_{\gamma \in \Gamma} \|q_1 - (q_2 \circ \gamma)\sqrt{\dot{\gamma}}\|,
\label{eq:d_a}
\end{equation} where
\(q(t) = \mbox{sign}(\dot{f}(t)) \sqrt{ |\dot{f}(t)|}\) is known as the
square-root slope function (SRSF) (\(\dot{f}\) is the time derivative
of \(f\)). The optimization problem in Equation \ref{eq:d_a} is most
commonly solved using a Dynamic Programming algorithm; see
\cite{robinson-2012} for a detailed description. If \(f\) is absolutely
continuous, then \(q\in\ltwo([0,1],\real)\) \cite{robinson-2012},
henceforth denoted by \(\ltwo\). For properties of the SRSF and the
reason for its use in this setting, we refer the reader to
\cite{srivastava-kalseen-joshi-jermyn:11}, \cite{marron2015} and
\cite{LRK}. Moreover, it can be shown that for any
\(\gamma_1, \gamma_2 \in \Gamma\), we have
\(d_a(f_1 \circ \gamma_1, f_2 \circ \gamma_2) = d_a(f_1, f_2)\), i.e.,
the amplitude distance is invariant to function warping.

\hypertarget{simplifying-geometry-of-gamma}{%
\subsection{\texorpdfstring{Simplifying Geometry of
\(\Gamma\)}{Simplifying Geometry of \textbackslash Gamma}}\label{simplifying-geometry-of-gamma}}

The space of warping functions, \(\Gamma\), is an infinite-dimensional
nonlinear manifold, and therefore cannot be treated as a standard
Hilbert space. To overcome this problem, we will use tools from
differential geometry to perform statistical analyses and to model the
warping functions. The following framework was previously used in
various settings including (1) modeling re-parameterizations of curves
\cite{srivastava-jermyn-PAMI:09}, (2) putting prior distributions on
warping functions \cite{Kurtek17} and \cite{Lu17}, (3) studying
execution rates of human activities in videos \cite{ashok-srivastava-etal-TIP:09}, and many others. It is also very
closely related to the square-root representation of probability density
functions introduced by \cite{bhattacharya-43}, and later used for
various statistical tasks (see e.g., \cite{SK}, \cite{SBK} and \cite{ASK}).

\begin{figure}[!t]
\begin{center}
  \begin{tikzpicture}
  	\def\R{2.5} 
    \def\angEl{20} 
	\def\angAz{-55} 
	\def\angPhiOne{67} 
	\pgfmathsetmacro\H{\R*cos(\angEl)} 
	\tikzset{xyplane/.estyle={cm={cos(\angAz),sin(\angAz)*sin(\angEl),-sin(\angAz),
	                              cos(\angAz)*sin(\angEl),(0,\H)}}}
	\tikzset{dot/.style={circle,draw=red!90,fill=red!90,minimum size=3.5pt}}
	\tikzset{dot2/.style={circle,draw=blue!90,fill=blue!90,minimum size=3.5pt}}
	\LongitudePlane[xzplane]{\angEl}{\angAz}
	\LongitudePlane[pzplane]{\angEl}{\angPhiOne}
	\LatitudePlane[equator]{\angEl}{0}
	\node at (.15,\R-1) [dot] (J) {};
	\node at (.15,\R+.5) [dot2] (10) {};
	\draw[green!90,line width=1.2pt, shorten <=-3pt,shorten >=-3pt] (J) -- (10);
	\fill[ball color=gray!30] (0,0) circle (\R);
	\draw[xyplane,blue,line width=1.2pt] (-1.25*\R,-1*\R) rectangle (1.25*\R,1*\R);
	\coordinate (O) at (0,0);
	\coordinate (N) at (0,\H);
	\coordinate[label=above:$\psi_1$] (N1) at (-1.25,1.8+\H);
	\coordinate[label=above:$v_i$] (v) at (-3.25,.8+\H);
	\coordinate[label=above:$\psi_i$] (N2) at (-3.25,\H-1);
	\DrawLatitudeCircle[\R]{0};
	\DrawLatitudeCircle[\R]{15};
	\DrawLatitudeCircle[\R]{30};
	\DrawLatitudeCircle[\R]{45};
	\DrawLatitudeCircle[\R]{60};
	\DrawLatitudeCircle[\R]{75};
	\DrawLatitudeCircle[\R]{89};
	\DrawLatitudeCircle[\R]{-15};
	\DrawLatitudeCircle[\R]{-30};
	\DrawLatitudeCircle[\R]{-45};
	\DrawLongitudeCircle[\R]{160};
	\DrawLongitudeCircle[\R]{140};
	\DrawLongitudeCircle[\R]{120};
	\DrawLongitudeCircle[\R]{100};
	\DrawLongitudeCircle[\R]{80};
	\DrawLongitudeCircle[\R]{60};
	\DrawLongitudeCircle[\R]{40};
	\DrawLongitudeCircle[\R]{20};
	\node at (2,\R) [dot2] (1) {};
	\node at (.8,\R-.5) [dot2] (2) {};
	\node at (1.3,1.6) [dot2] (3) {};
	\node at (-1.4,\R+.59) [dot2] (4) {};
	\node at (-2.1,\R+.12) [dot2] (5) {};
	\node at (-1,\R-.22) [dot2] (6) {};
	\node at (.9,\R+.15) [dot2] (7) {};
	\node at (.3,\R+.1) [dot2] (8) {};
	\node at (.2,1.8) [dot2] (9) {};
	\node at (.3,\R-.05) [dot] (A) {};
	\node at (.9,.55) [dot] (B) {};
	\node at (1.8,1.6) [dot] (C) {};
	\node at (-1.3,2.1) (D) {};
	\node at (-2,1) [dot] (E) {};
	\node at (.8,1.7) [dot] (F) {};
	\node at (-1,2) [dot] (G) {};
	\node at (.85,\R-.25) [dot] (H) {};
	\node at (.2,1.3) [dot] (I) {};
	\draw[-latex,blue!70,thick] (N1) -- (N);
	\draw[-latex,blue!70,thick] (v) -- (5);
	\draw[-latex,blue!70,thick] (N2) -- (E);
	\draw[green!90,line width=1.2pt, shorten <=-3pt,shorten >=-3pt] (B) .. controls (1.5,1.4) .. (3);
	\draw[green!90,line width=1.2pt, shorten <=-3pt,shorten >=-3pt] (C) .. controls (\R-.3,2.3) .. (1);
	\draw[green!90,line width=1.2pt, shorten <=-3pt,shorten >=-3pt] (D) .. controls (-1.5,\R+.5) .. (4);
	\draw[green!90,line width=1.2pt, shorten <=-3pt,shorten >=-3pt] (E) .. controls (-2.3,\R-.1) .. (5);
	\draw[green!90,line width=1.2pt] (A) -- (8);
	\draw[green!90,line width=1.2pt, shorten <=-3pt,shorten >=-3pt] (F) -- (2);
	\draw[green!90,line width=1.2pt, shorten <=-3pt,shorten >=-3pt] (G) -- (6);
	\draw[green!90,line width=1.2pt, shorten <=-3pt,shorten >=-3pt] (H) -- (7);
	\draw[green!90,line width=1.2pt, shorten <=-3pt,shorten >=-3pt] (I) -- (9);
\end{tikzpicture}
  \caption{Depiction of the SRSF space of warping functions as a sphere and a tangent space at $\psi_{1}$.}
  \label{fig:sphere-map}
\end{center}
\end{figure}
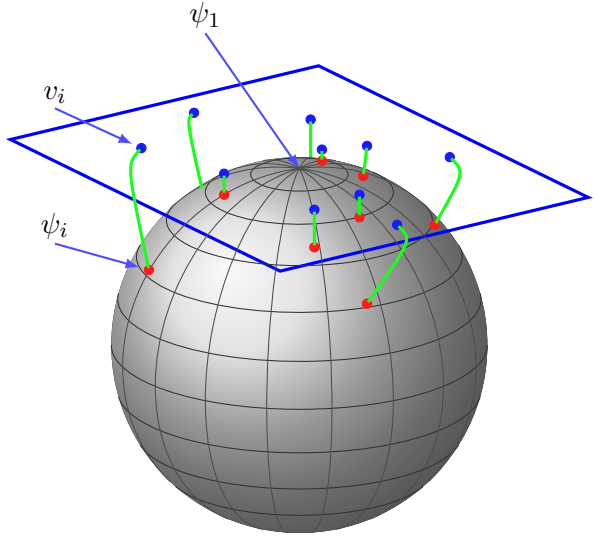

We represent an element \(\gamma \in \Gamma\) by the square-root of its
derivative \(\psi = \sqrt{\dot{\gamma}}\). Note that this is the same as
the SRSF defined earlier, and takes this form since
\(\dot{\gamma} > 0\). The identity \(\gamma_{id}\) maps to a constant
function with value \(\psi_{id}(t) = 1\). Since \(\gamma(0) = 0\), the
mapping from \(\gamma\) to \(\psi\) is a bijection and one can
reconstruct \(\gamma\) from \(\psi\) using
\(\gamma(t) = \int_0^t \psi(s)^2 ds\). An important advantage of this
transformation is that since
\(\| \psi\|^2 = \int_0^1 \psi(t)^2 dt = \int_0^1 \dot{\gamma}(t) dt = \gamma(1) - \gamma(0) = 1\),
the set of all such \(\psi\)s is the positive orthant of the unit Hilbert
sphere in \(\ltwo\): \(\Psi=\s_{\infty}^+\). In other words, the square-root representation simplifies
the complicated geometry of \(\Gamma\) to a (subset of a) unit sphere. The distance
between any two warping functions, i.e., the phase distance, is exactly
the arc-length between their corresponding SRSFs on \(\Psi\):
\begin{equation}
  d_{p}(\gamma_1, \gamma_2) = d_{\psi}(\psi_1, \psi_2) \equiv \cos^{-1}\left(\int_0^1 \psi_1(t) \psi_2(t) dt \right)\ .
\end{equation}
Figure \ref{fig:sphere-map} depicts the SRSF space
of warping functions as a unit sphere \cite{tucker-2014}.

\hypertarget{mapping-to-the-tangent-space-at-identity-element}{%
\subsection{Mapping to the Tangent Space at Identity
Element}\label{mapping-to-the-tangent-space-at-identity-element}}

While the geometry of \(\Psi\subset\s_{\infty}\) is more tractable, it
is still a nonlinear manifold and computing standard statistics remains
difficult. Instead, we use a tangent (vector) space at a certain fixed
point for further analysis. The tangent space at any point
\(\psi \in \Psi\) is given by:
\(T_{\psi}(\Psi) = \{v \in \ltwo| \int_0^1 v(t) \psi(t) dt = 0\}\). To
map between the representation space \(\Psi\) and tangent spaces, one
requires the exponential and inverse-exponential mappings. The
exponential map at a point \(\psi\in\Psi\) denoted by
\(\exp_\psi : T_{\psi}(\Psi) \mapsto \Psi\), is defined as

\begin{equation}
  \exp_\psi(v) = \cos(\|v\|)\psi+\sin(\|v\|)\frac{v}{\|v\|},
  \end{equation} \noindent where \(v\in T_{\psi}(\Psi)\). Thus,
\(\exp_\psi(v)\) maps points from the tangent space at \(\psi\) to the
representation space \(\Psi\). Similarly, the inverse-exponential map,
denoted by \(\exp_{\psi}^{-1} : \Psi \mapsto T_{\psi}(\Psi)\), is
defined as

\begin{equation}
  \exp_{\psi}^{-1}(\psi_1) = \frac{\theta}{\sin(\theta)}(\psi_1-\cos(\theta)\psi),
  \end{equation} \noindent where \(\theta = d_{p}(\gamma_1, \gamma)\).
This mapping takes points from the representation space to the tangent
space at \(\psi\).

The tangent space representation \(v\) is sometimes referred to as a
\emph{shooting vector}, as depicted in Figure \ref{fig:sphere-map}. The
remaining question is which tangent space should be used to represent
the warping functions. A sensible point on \(\Psi\) to define the
tangent space is at the sample Karcher mean \(\hat{\mu}_{\psi}\)
(corresponding to \(\hat{\mu}_{\gamma}\)) of the given warping
functions. For details on the definition of the sample Karcher mean and
how to compute it, please refer to \cite{tucker-wu-srivastava:2013}.

\hypertarget{model-for-combined-functional-principal-components}{%
\subsection{Model for Combined Functional Principal
Components}\label{model-for-combined-functional-principal-components}}

To model the association between the amplitude of a function and its
phase, Lee and Jung \cite{Lee:2017} use a concatenated function \(g^C\) on the
extended domain \([0,2]\) (for some \(C>0\)). The domain is extended to accommodate a combination of the
aligned (amplitude) function and the warping (phase) function. The motivation for the
combined function is to be able to properly handle properly the correlation
between the phase and amplitude components in functional data. Since the domain is \([0,1]\) for both, Lee and Jung \cite{Lee:2017} define the function \(g^C\) on the extended domain as follows:

\begin{equation}
  g^C(t) = \left\{\begin{array}{l r} f^*(t), & t\in[0,1) \\ C v(t-1), & t\in[1,2],\end{array}\right.
\end{equation} \noindent where \(f^*\) only contains the function's
amplitude (i.e., after groupwise alignment to the mean via SRSFs).
Furthermore, they assume that \(g^C\in \ltwo([0,2],\real)\). The
parameter \(C\) is introduced to adjust for the scaling imbalance
between \(f^*\) and \(v\). In the current work, we make a slight
modification to their method. In particular, it seems
more appropriate to construct the function \(g^C\) using the SRSF
\(q^*\) of the aligned function \(f^*\), since \(q^*\) is guaranteed to
be an element of \(\ltwo\). Thus, with a slight abuse in notation, we
proceed with the following joint representation of amplitude and phase:
\begin{equation}\label{eq:combfun}
  g^C(t) = \left\{\begin{array}{l r} q^*(t), & t\in[0,1) \\ C v(t-1), & t\in[1,2],\end{array}\right.
  \end{equation} where \(C\) is again used to adjust for the scaling
imbalance between \(q^*\) and \(v\).

Henceforth, we assume that \(q^*\) and \(v\) are both sampled using
\(T\) points, making the dimensionality of \(g^C\in\real^{2T}\). Then,
given a sample of amplitude-phase functions \(\{g^C_1,\dots,g^C_n\}\),
and their sample mean
\(\hat{\mu}_g^C = [\hat{\mu}_{q^*}~~ \hat{\mu}_v^C]\), we can compute
the sample covariance matrix as \begin{equation}
  K_g^C = \frac{1}{n-1}\sum_{i=1}^n (g^C_i - \hat{\mu}_g^C)(g^C_i - \hat{\mu}_g^C)^\T \in \mathbb{R}^{(2T) \times (2T)}\ .
\end{equation} The Singular Value Decomposition \(K_g^C=U_g^C\Sigma_g^C (U_g^C)^\T\) provides the joint principal
directions of variability in the given amplitude-phase functions as
the first \(p\leq n\) columns of \(U_g^C\). These can be converted back
to the original representation spaces (\(\mathcal{F}\) and \(\gamma\))
using the mappings defined earlier. Moreover, one can calculate the
observed principal coefficients as \(\inner{g^C_i}{U^C_{g,j}}\), for the
\(i^{th}\) function and the \(j^{th}\) principal direction of
variability. The superscript of \(C\) is used to denote the dependence
of the principal coefficients on the scaling factor.

This framework can be used to visualize the joint principal geodesic
paths. First, the matrix \(U^C_g\) is partitioned into the pair
\((U^C_{q^*},U^C_v)\). Then, the amplitude and phase paths within one
standard deviation of the mean are computed as

\begin{eqnarray}
  q^{*C}_{\tau,j} &=& \hat{\mu}_{q^*} + \tau \sqrt{\Sigma^C_{g,jj}}U^C_{q^*,j}, \label{eq:q} \\
  v^C_{\tau,j} &=& \tau \frac{\sqrt{\Sigma^C_{g,jj}}}{C} U^C_{v,j}~,
  \label{eq:v}
  \end{eqnarray} \noindent where \(\tau\in [-1,1]\), \(\Sigma_{g,jj}\)
and \(U^C_{j}\) are the \(j^{th}\) principal component variance and
direction of variability, respectively (note that the mean
\(\hat{\mu}_v^C\) is always zero). Then, one can obtain a joint
amplitude-phase principal path by composing \(f^{*C}_{\tau,j}\) (this is
the function corresponding to SRSF \(q^{*C}_{\tau,j}\)) with
\(\gamma^C_{\tau,j}\) (this is the warping function corresponding to
\(v^C_{\tau,j}\)).

The results of the above procedure clearly differ for variations of
\(C\). For example, using small values of \(C\), the first few principal
directions of variability will capture more amplitude variation, while
for large values of \(C\), the leading directions reflect more phase
variation. Lee and Jung \cite{Lee:2017} present a data-driven method for estimating \(C\) for a
given sample of functions. We use this approach in the current work to
determine an appropriate value of \(C\). Other approaches to choosing the value of \(C\) include (1) cross-validation metrics such as prediction performance, or (2) manual tuning based on which variability the user wants to emphasize in the statistical analysis.

\hypertarget{model}{%
\subsection{Statistical Model of Functions via fPCA}\label{model}}

There are several possibilities to develop statistical models for capturing the phase and amplitude
variability in functional data. Once we have obtained the
fPCA coefficients for the combined phase and amplitude variability we
can impose probability models directly on the coefficients. This in turn
induces a distribution on the function space \(\F\). Let
\(c = (c_1, \dots, c_{k})\) be the \(k\) dominant principal coefficients
of the combined model as described in the previous two sections. Recall
that the coefficients are constructed using
\(c_j = \inner{g^C}{U_{g,j}^C}\). The number $k$ is determined by the user and can be selected in different ways: (1) by minimizing cross-validated fPCA reconstruction error, (2) by retaining the smallest number of fPCA coefficients that explain at least $X\%$ of the variability in the given data ($X$ is usually chosen as a large number, e.g., 90 or 95), or (3) by iteratively comparing tolerance bounds using an increasing number of fPCA components and stopping when the bounds change negligibly. In our approach, for simplicity, we use method (2) with $X=90$.

The vector \(c\) is modeled using a multivariate Gaussian probability
distribution with zero mean and covariance \(\Sigma\), i.e., \( c\sim\mathcal{N}_k(0, \Sigma) \). By construction of the principal coefficients, the mean vector is
zero and the covariance is a \(k\times k\) diagonal matrix. The diagonal
elements of the covariance are estimated directly using the eigenvalues
of the sample covariance matrix, \(\hat{\sigma}_1^C,\dots,\hat{\sigma}_k^C\). The
model on the fPCA scores induces a probability model on \(\F\) and
provides a means of efficiently sampling functions that exhibit the
amplitude and phase variability of the original data.

\hypertarget{functional-tolerance-bounds}{%
\section{Functional Tolerance
Bounds}\label{functional-tolerance-bounds}}

In this section, we provide two methods for calculating tolerance bounds
for functional data in the presence of warping variability. In general,
we seek bounds which guarantee that with \((1-\alpha)100\%\) confidence,
\((1-p)100\%\) of the data falls within these bounds \citep{Hahn:2011}.
Our first approach uses the parametric bootstrap to construct tolerance
bounds sampling from the model on the fPCA coefficients. The second
method provides the tolerance bound in the fPCA coefficient space using
a tolerance factor based on the multivariate Gaussian model. We provide
a detailed description of both of these procedures next.

\hypertarget{boot}{%
\subsection{Method 1: Bootstrapped Geometric Tolerance
Bounds}\label{boot}}

First, we construct statistical bounds using bootstrapping from the
fPCA-based model described in Section \ref{model}, and provide a means
of characterizing the uncertainty in the original functional data.
Bootstrapping refers to repeated sampling from the model, and this
process can be used to construct confidence bounds for essentially any
quantity of interest. For a detailed overview of statistical bootstrap
techniques see \cite{davison:1997} and \cite{davison:2003}.

For two-sided tolerance bounds, there is both an upper and a lower
bound. The upper tolerance bound is an upper confidence bound on an
upper population quantile. The lower tolerance bound is a lower
confidence bound on a lower population quantile. In this sense,
tolerance bounds are simply confidence bounds on population quantiles.
Construction of equal-tailed tolerance bounds using the bootstrap
approach is described as the following procedure:

\begin{enumerate}
\def\labelenumi{\arabic{enumi}.}
\tightlist
\item
  Sample \(n\) functions from the constructed generative model described
  in Section \ref{model}, resulting in a random sample
  \(g^C_i,~i=1,\dots,n\).
\item
  From \(g^C_i\) extract the amplitude functions (SRSFs) \(q_i^*\) and
  vectors \(v_i\). The random warping functions can be constructed using
  \(\gamma_i(t) = \int_0^t(\exp_{\hat{\mu}_\psi}(v_i(s)))^2ds\).
\item
  Estimate the \(p/2\) and \((1-p/2)\) quantiles of the set of random
  SRSF-based amplitudes and random warping functions, denoted by
  \((q^*_{p/2},q^*_{1-p/2})\) and \((\gamma_{p/2},\gamma_{1-p/2})\),
  respectively. There are many ways to estimate quantiles for functional
  data, with the cross-sectional (pointwise) approach being most popular
  \cite{Lewis:2017}. We propose to use the geometric method of
  Xie et al. \cite{XieKurtek:2016}, which relies on the Riemannian geometry of the
  amplitude and phase spaces. In that paper, the authors only compute
  quartiles, but their method can be easily extended to calculate
  general quantiles.
\item
  Repeat steps 1-3 \(S\) times for a large \(S\) (\(S\) should be large
  enough relative to \(\alpha\) to provide stable bounds). This results
  in a collection (of size \(S\)) of \((q^*_{p/2},q^*_{1-p/2})\) and
  \((\gamma_{p/2},\gamma_{1-p/2})\).
\item
  Calculate the \(\alpha/2\) and \((1-\alpha/2)\) quantiles of the \(S\)
  samples of \((q^*_{p/2},q^*_{1-p/2})\) and
  \((\gamma_{p/2},\gamma_{1-p/2})\). These quantiles form
  \((1-\alpha)100\%\) bootstrap tolerance bounds with \((1-p)100\%\)
  coverage.
\item
  The amplitude tolerance bounds are pulled back to the quotient space
  \(\F/\Gamma\) via integration. For display of the quantiles we use
  the surface method of Xie et al. \cite{XieKurtek:2016} for both the amplitude
  and phase components.
\end{enumerate}

\hypertarget{pcatb}{%
\subsection{Method 2: Combined fPCA-based Tolerance
Region}\label{pcatb}}

The second approach to construct tolerance bounds is to directly use the
fPCA-based multivariate Gaussian model that is imposed on the
principal coefficients. Then, one can construct multivariate tolerance
regions on the coefficient space using the model given in \cite{Krishnamoorthy:2006}. We provide more details of this approach next.

Let \(\x_1,\dots,\x_n\in\real^d\) be a
random sample from a \(k\)-variate Gaussian distribution with mean
vector \(\mu\) and covariance matrix \(\Sigma\). The sample mean vector
\(\bar{\x}\) and the sums of squares and cross-product matrix \(A\) are
defined as: \begin{eqnarray} \nonumber
\bar{\x} &=& \frac{1}{n}\sum_{i=1}^n \x_i,\quad
A = \sum_{i=1}^n (\x_i-\bar{\x})(\x_i-\bar{\x})^\T. \nonumber
\end{eqnarray} A tolerance region that contains at least \(p\)
proportion of the data from a \(\mathcal{N}_k(\mu,\Sigma)\) distribution
with \(\beta\) confidence is given by \begin{equation}
\{\x:(n-1)(\x-\bar{\x})^\T A^{-1}(\x-\bar{\x})\leq b\}\label{eq:tolfact}
\end{equation} The parameter \(b\) is known as the \emph{tolerance
factor}, and is determined by the probability condition $P_{\bar{\x},A}(P_\x((n-1)(\x-\bar{\x})^\T A^{-1}(\x-\bar{\x})\leq b|\bar{\x},A)\geq p) = \beta$. The exact method of computing \(b\) is known to be extremely
difficult and there are multiple approximations that have been proposed
in the literature (see \cite{Krishnamoorthy:2009} for multiple
methods). In this work, we use the approach of Krishnamoorthy and Mondal
\cite{Krishnamoorthy:2006} due to its known accuracy and precision.
Note that in the fPCA coefficient space \(\bar{x} = 0\) by construction,
which simplifies the computation. Once the tolerance factor is computed, a new sample function can be tested against this factor using Equation \ref{eq:tolfact}.

\hypertarget{simulations}{%
\section{Simulation Results}\label{simulations}}

In this section, we present results on a simulated dataset and compare
the proposed approaches to the recent methods of Lewis et al. \cite{Lewis:2017} and
Rathnayake and Choudhary \cite{Rathnayake:2016}. The method of Lewis et al. is most closely related to
our work and is thus important to compare to. In all of the plots in
the following sections, we re-scale the domain of the functional
observations and warping functions to \([0,1]\) for simplicity.

\hypertarget{bootstrapped}{%
\subsection{Method 1: Bootstrapped Geometric Tolerance
Bounds}\label{bootstrapped}}

\begin{figure}[!t]
{\centering \includegraphics[width=\textwidth]{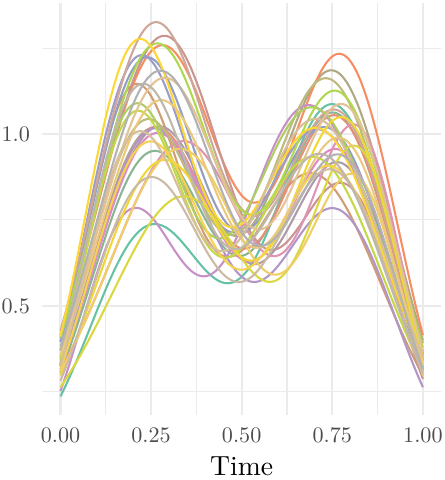} 
}
\caption{Alignment of the simulated data set. (a) Original functions. (b) Aligned functions (amplitude). (c) Warping functions (phase).}\label{fig:simu_data}
\end{figure}

First, we provide a numerical simulation for the bootstrapped geometric
tolerance bounds method. For this purpose, we generate data previously
used in \cite{kneip-ramsay:2008}. The individual functions are given
by:
\[y_i(t) = z_{i,1} e^{-(t-1.5)^2/2} + z_{i,2}e^{-(t+1.5)^2/2},\ t \in [-3, 3],\ i=1,2,\dots, 21,\]where
\(z_{i,1}\) and \(z_{i,2}\) are \emph{i.i.d.}
\(\mathcal{N}(1, (0.25)^2)\). Each of the simulated functions are then
warped according to:
\[\gamma_i(t) = 6\left(\frac{e^{a_i(t+3)/6} -1}{e^{a_i} - 1}\right) - 3,\ \text{if}\ a_i \neq 0,\ \text{otherwise}\ \gamma_i = \gamma_{id},\]where
\(\gamma_{id}(t) = t\) is the identity warping. Here, \(a_i\) are
equally spaced between \(-1\) and \(1\), and the observed functions are
computed by composition using \(f_i = y_i \circ \gamma_i\). A set of 30
such functions forms the original data and is shown in Figure
\ref{fig:simu_data}(a). The aligned functions (amplitude) and
corresponding warping functions (phase) are shown in Figure
\ref{fig:simu_data}(b) and (c), respectively.

\begin{figure}[!t]

{\centering \subfloat[1st PD\label{fig:pca_plot1}]{\includegraphics{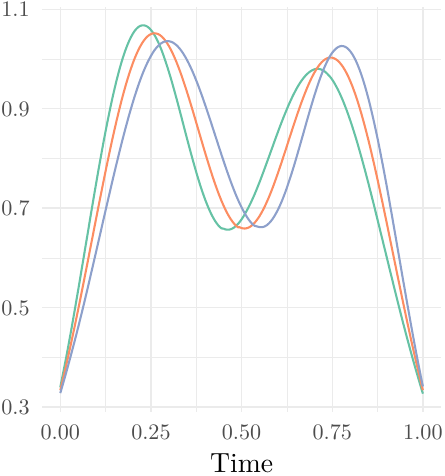} }\subfloat[2nd PD\label{fig:pca_plot2}]{\includegraphics{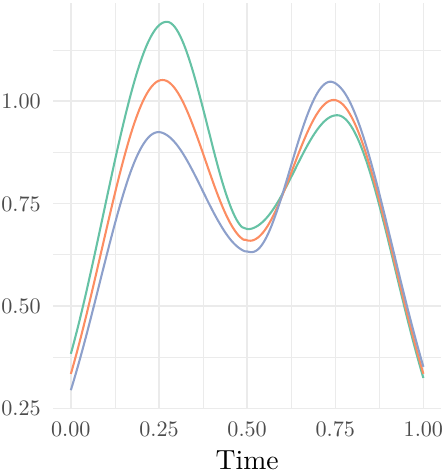} }\subfloat[3rd PD\label{fig:pca_plot3}]{\includegraphics{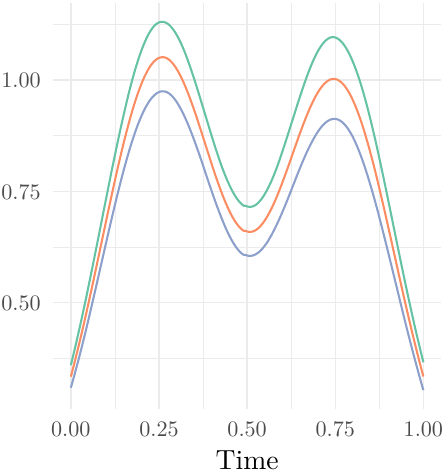} }

}

\caption{Joint principal directions (PD) of variability for the simulated dataset for $\tau=-1$ (blue), $0$ (red), $1$ (green).}\label{fig:pca_plot}
\end{figure}

Given that we have the phase-amplitude separation of the simulated data,
we can calculate the principal directions of variability using the
combined fPCA method defined in Section \ref{pca}. Figure
\ref{fig:pca_plot} shows the results of this procedure on the simulated
dataset presented in Figure \ref{fig:simu_data}. We plot the joint
amplitude-phase principal geodesic paths using Equations \eqref{eq:q} and
\eqref{eq:v}, for \(\tau= -1, 0, 1\) and \(j = 1, 2, 3\). The first
three singular values for this data are 4.97, 4.24 and 0.77, with the
rest being fairly small. The first four principal directions of variability accounted for 90\% of the
overall variability in the data. The first principal direction of variability displayed in
panel (a) mostly corresponds to phase variation, while the second direction in panel (b) captures the height (amplitude) variation of the
functions. The third principal direction exhibits very small scaling variability in panel (c).

Next, we calculate tolerance bounds for the simulated data using the
bootstrap approach presented in Section \ref{boot}. Using the combined
amplitude and phase fPCA computed in the previous section, we impose a
multivariate Gaussian model using the first four principal directions.
We perform 500 bootstrap re-samples; within each iteration, we sample 30
functions to calculate the tolerance bounds. We use this procedure to
compute the tolerance bound with 99\% coverage with a confidence level
of 95\%.

\begin{figure}

{\centering \subfloat[Amplitude\label{fig:boot_amp1}]{\includegraphics{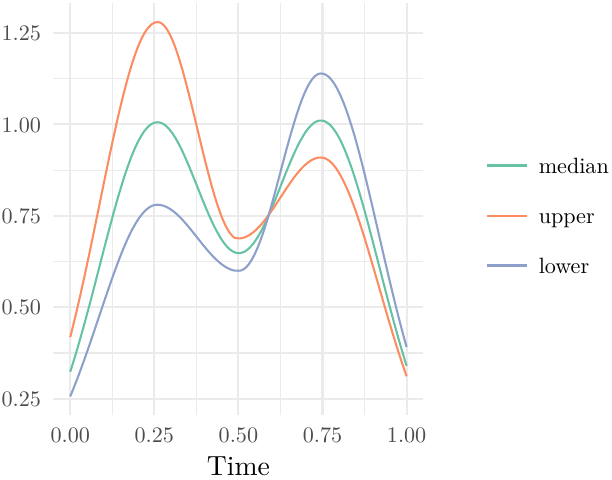} }\subfloat[Phase\label{fig:boot_amp2}]{\includegraphics{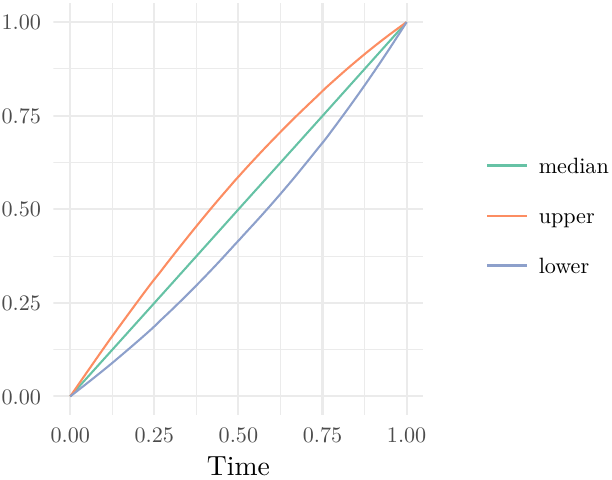} }

}

\caption{Bootstrapped geometric tolerance bounds for the simulated dataset.}\label{fig:boot_amp}
\end{figure}

Figure \ref{fig:boot_amp} presents the tolerance bounds for the (a)
amplitude and (b) phase. The upper bound is shown in red and the lower
bound is shown in blue (the median is also plotted in green). Given the structure of the original data, the
bounds are intuitive for the phase component. However, the amplitude
bounds overlap, i.e., for the first peak the lower bound is above the
upper bound while for the second peak, the lower bound is below the
upper bound. While this result may seem counterintuitive at first, it
stems from the geometric approach to generate the quantiles. In
particular, the amplitude component captures the ``shape'' of the
functions, and the quantiles are constructed directly on the amplitude
space. Thus, it is more appropriate to think of the bounds as points on
that space rather than pointwise (cross-sectional) bounds as currently
done in the literature. As a result, there is no guarantee that the
bounds will not overlap as they are constructed to capture the global
``shape" tolerance region on the amplitude function space. This will
also be seen in the real data applications.

\begin{figure}[!t]

{\centering \subfloat[Amplitude\label{fig:boot_surf_amp1}]{\includegraphics{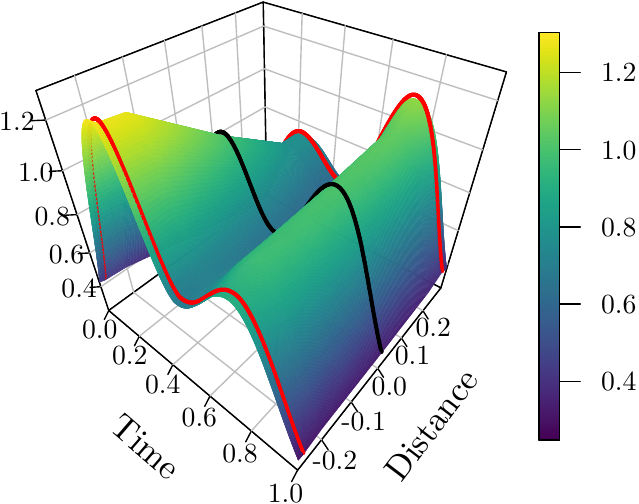} }\subfloat[Phase\label{fig:boot_surf_amp2}]{\includegraphics{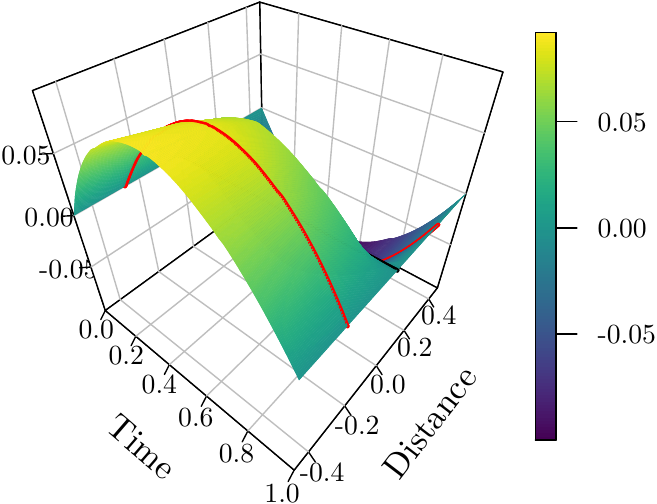} }

}

\caption{Surface plots of the bootstrapped geometric tolerance bounds for the simulated dataset.}\label{fig:boot_surf_amp}
\end{figure}

Xie et al.~\citep{XieKurtek:2016} provide a detailed commentary of this in
their paper and propose a surface plot using the proper metrics to
display the quantiles. We thus use the same approach here, and present
such surface plots for the amplitude and phase components in Figure
\ref{fig:boot_surf_amp}(a) and (b), respectively, with the bounds shown
in red. In panel (a), the amplitude component of the median and bounds
is shown along the \(z\)-axis. Furthermore, the functions are separated
according to the pairwise amplitude distances between them. Based on
this surface plot we now have a natural view of the amplitude
variability in the data, with the bounds separated by appropriate
distances for a clear display. Panel (b) provides a similar display for
the phase component. In this case, it is more effective to display the
difference between the median/bounds and the identity element
\(\gamma_{id}\). Thus, as expected, the median is very close to the
constant function 0. The red warping bounds reflect natural variability
in the original data and are approximately equidistant from the median.

\begin{figure}[!t]

{\centering \subfloat[Rathnayake and Choudhary \cite{Rathnayake:2016}\label{fig:lewis_boot_amp1}]{\includegraphics{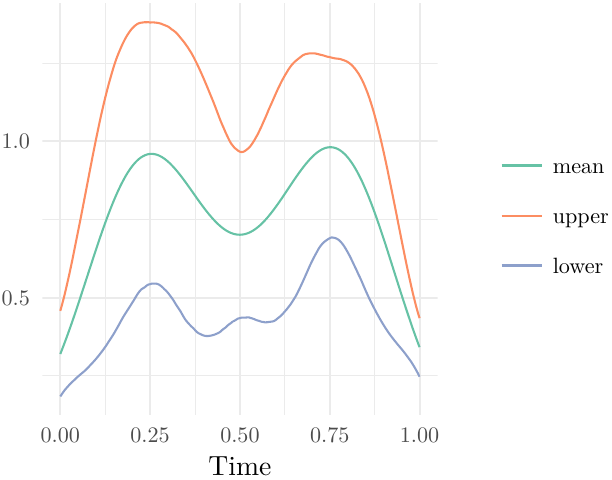} }\subfloat[Lewis et al. \cite{Lewis:2017}\label{fig:lewis_boot_amp2}]{\includegraphics{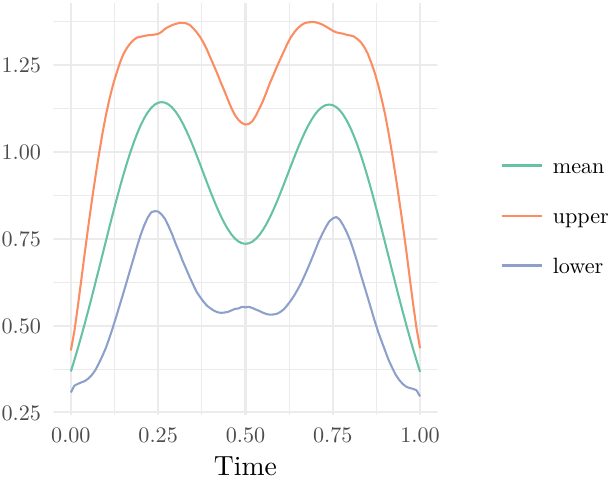} }

}

\caption{Tolerance bounds for the simulated dataset constructed using the methods in \cite{Rathnayake:2016} and \cite{Lewis:2017}.}\label{fig:lewis_boot_amp}
\end{figure}

Figure \ref{fig:lewis_boot_amp}(a) presents the tolerance bounds
calculated using the approach of Rathnayake and Choudhary \cite{Rathnayake:2016}. Comparing
these bounds with those presented in Figure \ref{fig:boot_amp}, we see
some striking structural differences. In particular, in Figure
\ref{fig:lewis_boot_amp}(a), it is difficult to determine the relative
contributions of amplitude and phase to the tolerance bounds. Moreover,
the upper and lower bounds are not representative of the actual shape of
the original functions, and both exhibit more than just the two peaks
found in the data. This is due to the fact that the bounds were computed
in a cross-sectional manner without accounting for warping variability. Figure \ref{fig:lewis_boot_amp}(b) presents the tolerance bounds
calculated using the approach of Lewis et al. \cite{Lewis:2017}. Again, comparing these
bounds with those presented in Figure \ref{fig:boot_amp}, we see some
structural differences. Both the upper and lower bounds do not
accurately capture the true underlying shape of the given data.

\begin{table}[!t]
\centering
\begin{tabular}{rrrr}
  \hline
 & 99\% & 95\% & 90\% \\
  \hline
Amplitude & 100.00 & 97.60 & 93.80 \\
  Phase & 99.80 & 98.40 & 90.00 \\
  \hline
  Rathnayake and Choudhary \cite{Rathnayake:2016} & 75.00 & 70.00 & 67.00\\
  Lewis et al. \cite{Lewis:2017} & 71.00 & 67.00 & 64.00\\
   \hline
\end{tabular}
\caption{Simulated confidence values of 90\% coverage tolerance using the bootstrap-based approach (top). Corresponding estimated confidence values using the methods in \cite{Rathnayake:2016} and \cite{Lewis:2017}.}
\label{tab:simu_coverage}
\end{table}

\begin{figure}[!t]

{\centering \subfloat[Amplitude\label{fig:boot_surf_amp2a1}]{\includegraphics{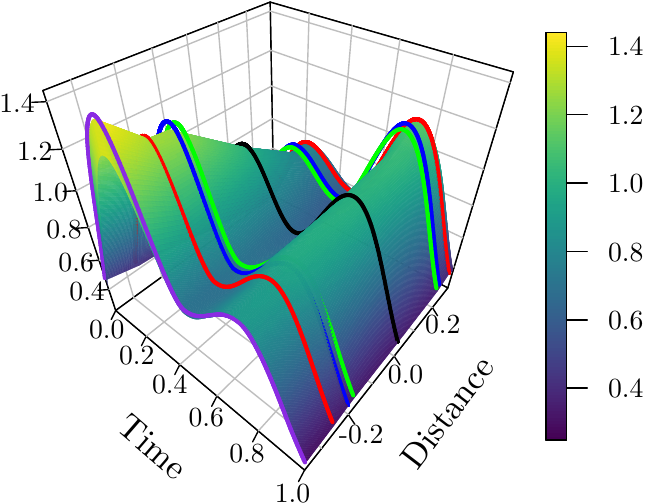} }\subfloat[Phase\label{fig:boot_surf_amp2a2}]{\includegraphics{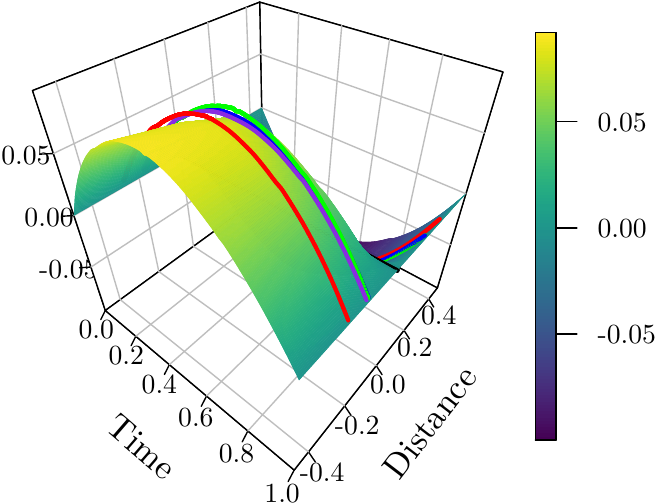} }

}

\caption{Surface plots of the bootstrapped geometric tolerance bounds for the simulated dataset for different confidence levels (90\%=green, 95\%=blue, 99\%=red). An outlier amplitude and phase function is shown in purple.}\label{fig:boot_surf_amp2a}
\end{figure}

Next, we generate an additional set of random functions from our
combined fPCA model to check the coverage of the calculated tolerance
bounds. We study the performance of our method based on three sets of
tolerance bounds at the 90\% coverage rate with confidence levels of
90\%, 95\% and 99\%. We first generate 100 random functions; for each
function, we compute its SRSF (\(q\)) and warp it to \(\hat{\mu}_q\) to
extract the corresponding warping function \(\gamma\) and the aligned
SRSF (\(q^*\)). We then compute the 90\% quantile, and compare \(q^*\)
and \(\gamma\) to the corresponding tolerance bounds (the entire
function must fall within the tolerance bounds). This process is repeated
500 times and the corresponding estimated confidence values are listed in
Table \ref{tab:simu_coverage} for each of the corresponding true confidence values. For both amplitude and phase tolerance bounds, the
calculated confidence values are slightly higher than expected; this
suggests that the bounds we compute are conservative.
Nonetheless, the simulated values are very close to the correct
confidence levels. For comparison, we also list the estimated confidence values computed using the methods defined in \cite{Rathnayake:2016} and \cite{Lewis:2017}. For both methods, the estimated confidence values are much lower than expected. This is most likely due to the fact that these approaches do not appropriately account for the amplitude and phase variabilities in the given data.

Next, we show the effect of the confidence level on the tolerance bounds.
Figure \ref{fig:boot_surf_amp2a} presents surface plots of the tolerance
bounds for (a) amplitude and (b) phase for three different settings. The red curves are the
99\% confidence tolerance bounds, the blue curves are the 95\% confidence
tolerance bounds, and the green curves are the 90\% confidence
tolerance bounds. All of these bounds were generated for 90\% coverage.
For both amplitude and phase, as the confidence level increases the bounds move outward in distance from the
median as one would expect. For the phase, the 95\%
and 90\% confidence tolerance bounds are extremely close. The purple amplitude function in panel (a) falls outside the amplitude tolerance bounds for all three confidence levels; in panel (b), its phase component also falls outside the 90\% confidence tolerance bounds for phase. The surface plots are able to accurately display the bounds by nicely separating them according to amplitude or phase distance.

\begin{figure}[!t]

{\centering \subfloat[Amplitude\label{fig:simu_pca_comp1}]{\includegraphics{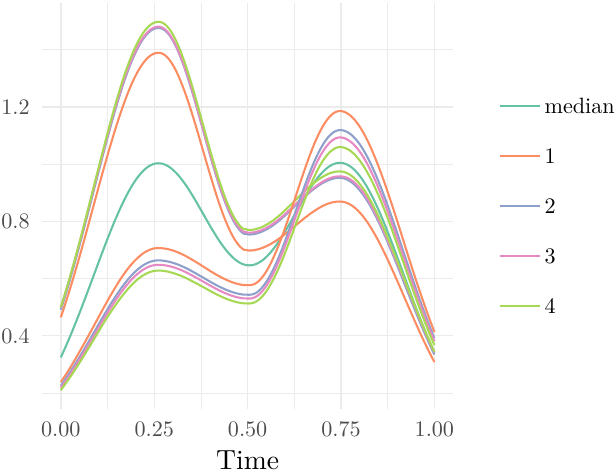} }\subfloat[Phase\label{fig:simu_pca_comp2}]{\includegraphics{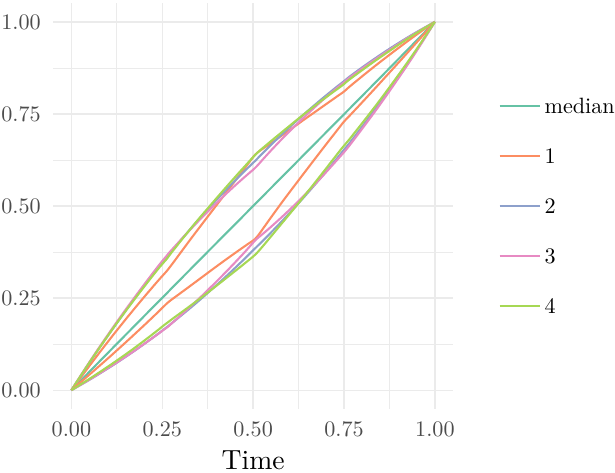} }

}

\caption{Effects of the number of principal directions of variability retained by the combined amplitude and phase fPCA model on the resulting
tolerance bounds for 99\% coverage with 95\% confidence.}\label{fig:simu_pca_comp}
\end{figure}

To show the effects of the number of principal directions of variability retained by the combined
amplitude and phase fPCA model on the resulting
tolerance bounds, we applied the bootstrap method
for 1-4 principal directions. The results are shown in Figure
\ref{fig:simu_pca_comp}. As the dimensionality of the model increases to 3 or 4, we see the bounds remain relatively unchanged showing good stability. With enough principal components (approximately four in this example), the constructed tolerance bounds are able to more accurately represent the variability found in the original data, for both phase and amplitude.

\begin{figure}[!t]

{\centering \includegraphics{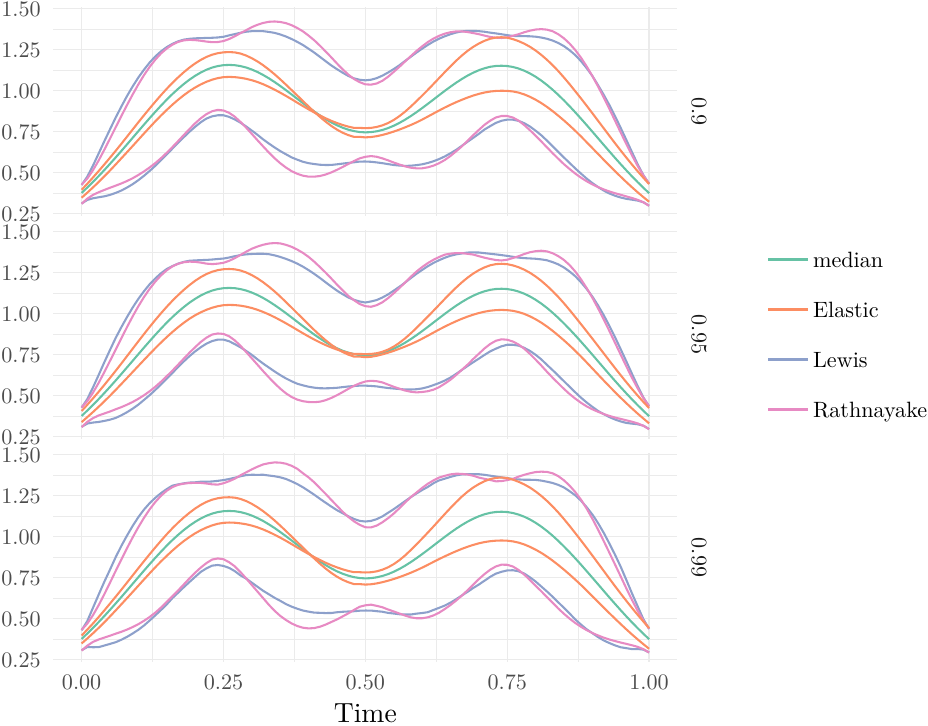}

}

\caption{Comparison of the proposed method (Elastic) to \cite{Rathnayake:2016} (Rathnayake) and \cite{Lewis:2017} (Lewis). We use the simulated data to construct tolerance bounds for confidence levels 90\% (top), 95\% (middle), and 99\% (bottom) at 99\% coverage.}\label{fig:simu_conf_comp}
\end{figure}

\begin{figure}[!t]

{\centering \includegraphics{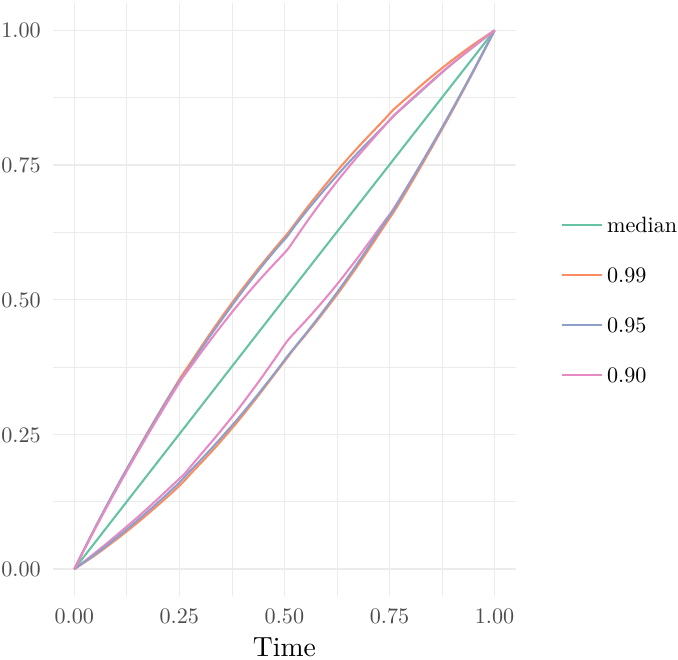}

}

\caption{Proposed phase tolerance bounds, constructed using the simulated data, for confidence levels 90\%, 95\% and 99\% at 99\% coverage.}\label{fig:simu_conf_comp_ph}
\end{figure}

Lastly, we compare the proposed tolerance bounds to those of Rathnayake and Choudhary \cite{Rathnayake:2016} and Lewis et al. \cite{Lewis:2017} for different confidence levels at 99\% coverage. These results are presented in Figure \ref{fig:simu_conf_comp} for 90\%
(top), 95\% (middle) and 99\% (bottom) confidence levels. The proposed approach (labeled Elastic) is able to
accurately capture the shape of the given data in the tolerance bounds for all confidence levels; we also note that, overall, the bounds spread outward as the confidence level increases. As seen previously, both of the competing methods (labeled Rathnayake for \cite{Rathnayake:2016} and Lewis for \cite{Lewis:2017}) produce multimodal tolerance bounds, where the various modes increase or decrease as the confidence level changes.
This comparison was performed only on the amplitude, as the methods in \cite{Rathnayake:2016} and \cite{Lewis:2017} do not consider phase and
amplitude separately. Figure \ref{fig:simu_conf_comp_ph} provides the phase tolerance bounds for the same varying levels of confidence at 99\% coverage for the proposed method.

\hypertarget{method-2-combined-fpca-based-tolerance-region}{%
\subsection{Method 2: Combined fPCA-based Tolerance
Region}\label{method-2-combined-fpca-based-tolerance-region}}

In this section, we calculate tolerance bounds for the simulated dataset
using the fPCA basis approach presented in Section \ref{pcatb}. Again,
we impose a multivariate Gaussian model using the first four principal
directions of variability. We calculate a
tolerance region with 99\% coverage with a confidence level of 95\%.
In contrast to the bootstrap approach described in the previous section, this method constructs the tolerance region directly on the
fPCA coefficient space.

\begin{figure}[!t]

{\centering \includegraphics{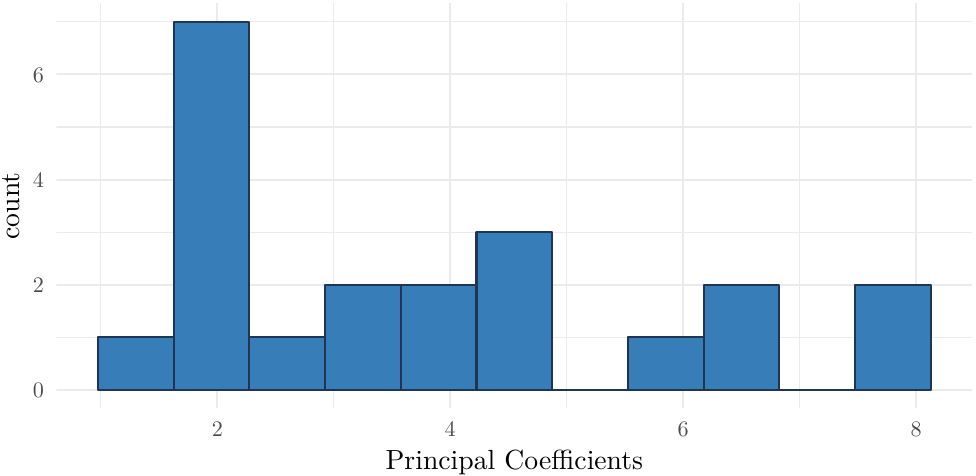}

}

\caption{Histogram of tolerance scores for each function in the simulated dataset after projection onto the first four fPCA basis functions.}\label{fig:pcatol2}
\end{figure}

First, we must calculate the tolerance factor, which is done
using Algorithm 2 in \cite{Krishnamoorthy:2006} with 100,000 iterations.
The resulting tolerance factor \(b\) for this dataset, and the retained
dimension of four, is 32.0027. For different dimensions of the
multivariate Gaussian model, the tolerance factor can be easily
calculated using the above-mentioned algorithm or via the tables
provided by \cite{Krishnamoorthy:2006}. Each new function that needs to be
tested can be projected onto the fPCA basis. Based on the resulting
fPCA coefficients, we compute the function's tolerance score using Equation \ref{eq:tolfact} and compare
it to the tolerance factor. Figure \ref{fig:pcatol2} presents a histogram of the tolerance scores
for all 21 functions in the simulated dataset. All of the scores fall
well-below the calculated tolerance factor; this is expected as the
tolerance bound has 99\% coverage, and was calculated using the same
data.

Again, we generate an additional set of random functions from our model
to check the coverage of the calculated tolerance region. As previously,
we study the performance of our method based on three sets of tolerance
regions at the 90\% coverage rate with confidence levels of 90\%, 95\%,
and 99\%. We first generate 100 random functions, and for each function,
we compute its SRSF (\(q\)). The SRSF is then warped to \(\hat{\mu}_q\)
to extract the corresponding warping function \(\gamma\) and aligned
SRSF (\(q^*\)). The warping function is mapped to the tangent space at
the warping mean, which can be computed using the algorithm described in
\cite{XieKurtek:2016}. The function \(g^C\) is then calculated and
projected into the fPCA coefficient space. We calculate the 90\% quantile in the
coefficient space as well as the tolerance scores for each of the
functions; the tolerance scores are then compared to the tolerance
factor. We repeat this process 500 times, and report the estimated
confidence values in Table \ref{tab:simu_coverage_pca} for each
tolerance bound. Similarly to the bootstrap method, the simulated confidence values
are conservative.

\begin{table}[!t]
\centering
\begin{tabular}{rrrr}
  \hline
 & 99\% & 95\% & 90\% \\
  \hline
fPCA & 99.20 & 98.40 & 97.00 \\
   \hline
\end{tabular}
\caption{Simulated confidence values of 90\% coverage tolerance using the combined fPCA-based approach.}
\label{tab:simu_coverage_pca}
\end{table}

\hypertarget{application-to-real-data}{%
\section{Applications to Real Data}\label{application-to-real-data}}

Here, we present results on two real datasets: (1) axial weld data and
(2) PQRST complexes extracted from electrocardiogram (ECG) signals \cite{kurtek:2013}. For each of the examples, we study the
effectiveness of the proposed approach in the calculation of the
tolerance bounds, specifically in capturing the amplitude and phase
variabilities.

\hypertarget{weld-data}{%
\subsection{Tolerance Bounds for Weld Data}\label{weld-data}}

\begin{figure}[!t]

{\centering \includegraphics[width=5.5in]{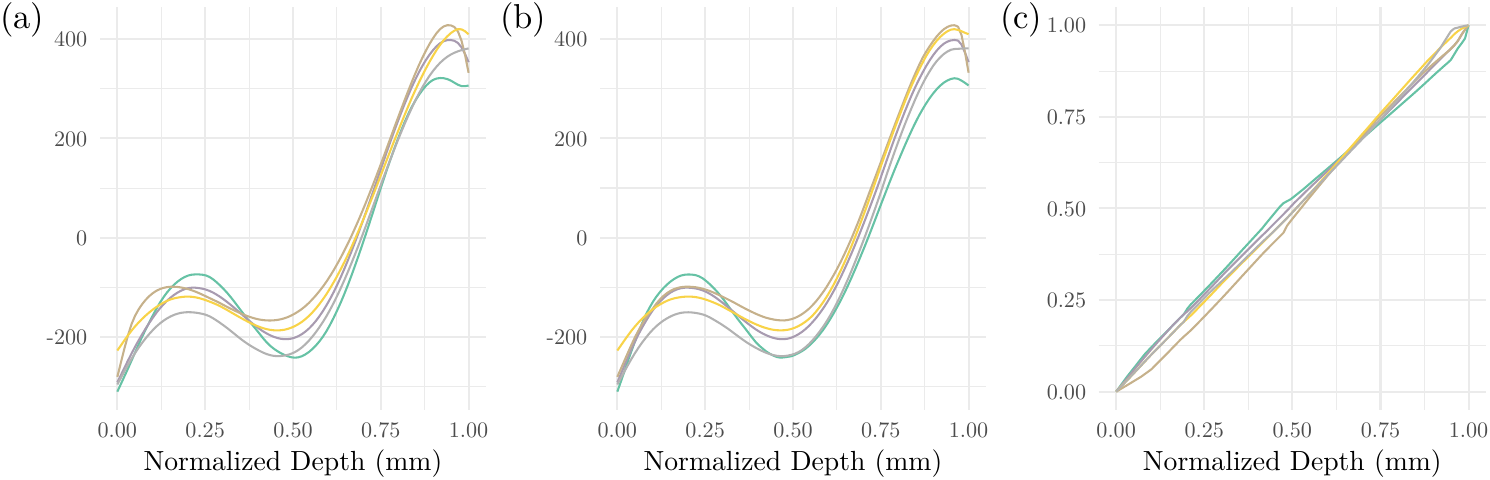}

}

\caption{Alignment of the WRS axial contour weld data. (a) Original functions. (b) Aligned functions (amplitude). (c) Warping functions (phase).}\label{fig:weld_data}
\end{figure}

Weld residual stress (WRS) is a main component crack formation in safety
critical pipe welds within nuclear power plants \cite{benson:2015}.
The stress cannot be measured directly and involves releasing strain
from a weld specimen and converting measured displacements to stress
using a mathematical model. Two measurement methods, both destructive to
the specimen, are the deep hole drilling method and the contour method
\cite{prime:2004}, \cite{mahmoudi:2008}. The measurements can help
assess the validity of WRS predictions for new welds \cite{Lewis:2017}.
Figure \ref{fig:weld_data}(a) shows original contour method measurements
along the axial direction of a weld for five different locations along
the circumference of the weld. The values represent stress as a function
of normalized weld depth. Though the phase variability is small relative
to the amplitude variability, it is important to account for both since
WRS predictions affect crack-growth calculations used in plant safety
assessments \cite{benson:2015}. With a small dataset, it is difficult to
assess the validity of tolerance bound coverage rates. However, our
motivation for tolerance bounds is to provide a preliminary estimate of
the uncertainty in the WRS measurements which would be assessed by
subject matter experts before applying them to crack-growth
calculations. Figure \ref{fig:weld_data}(b) and (c) show the aligned
functions (amplitude) and warping functions (phase), respectively. The
main source of phase variability in this data comes from the difference
in location of the first peak across measurements.

\begin{figure}

{\centering \subfloat[Amplitude\label{fig:weld_boot_amp1}]{\includegraphics{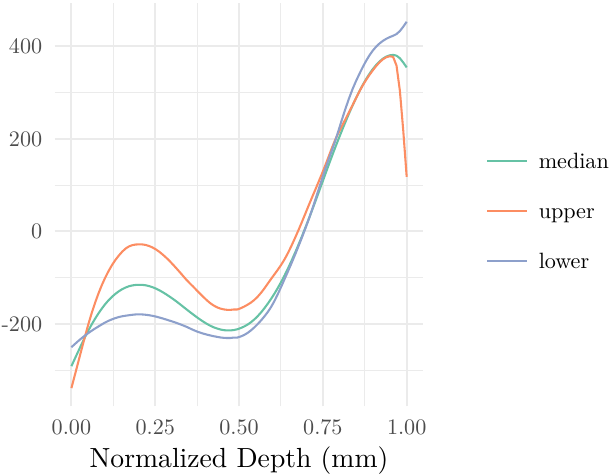} }\subfloat[Phase\label{fig:weld_boot_amp2}]{\includegraphics{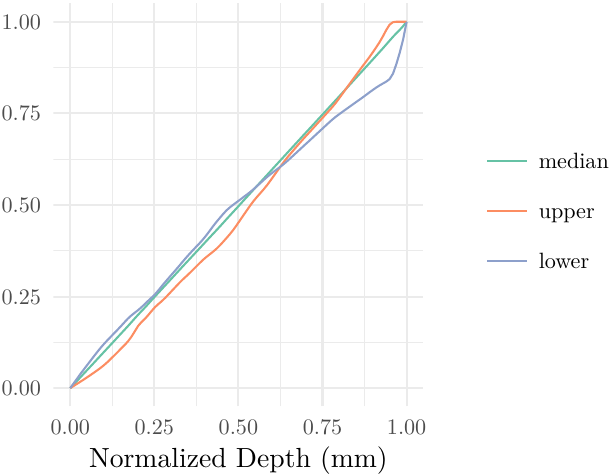} }

}

\caption{Bootstrapped geometric tolerance bounds for the WRS axial contour weld data.}\label{fig:weld_boot_amp}
\end{figure}

\begin{figure}[!t]

{\centering \subfloat[Amplitude\label{fig:weld_boot_surf_amp1}]{\includegraphics{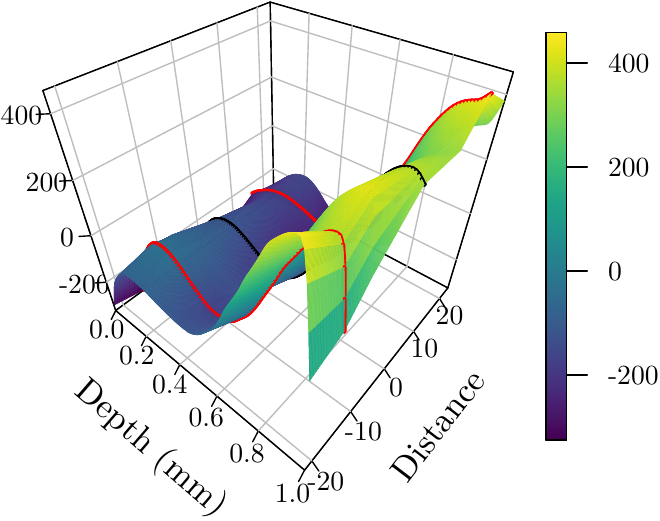} }\subfloat[Phase\label{fig:weld_boot_surf_amp2}]{\includegraphics{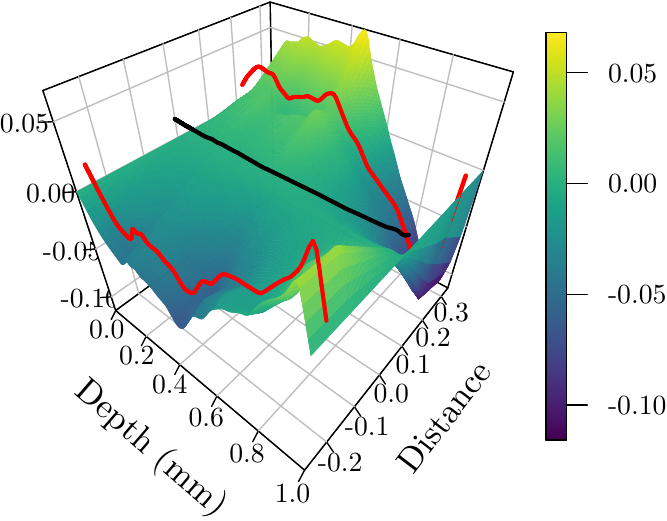} }

}

\caption{Surface plots of the bootstrapped geometric tolerance bounds for the WRS axial contour weld data.}\label{fig:weld_boot_surf_amp}
\end{figure}

\begin{figure}[!t]

{\centering \subfloat[Rathnayake and Choudhary \cite{Rathnayake:2016}\label{fig:weld_lewis_boot_amp1}]{\includegraphics{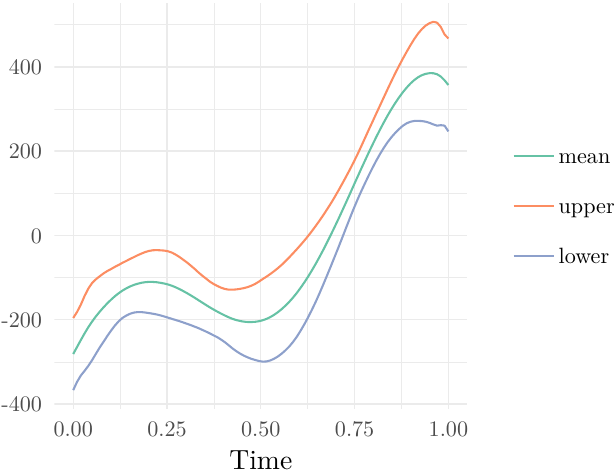} }\subfloat[Lewis et al. \cite{Lewis:2017}\label{fig:weld_lewis_boot_amp2}]{\includegraphics{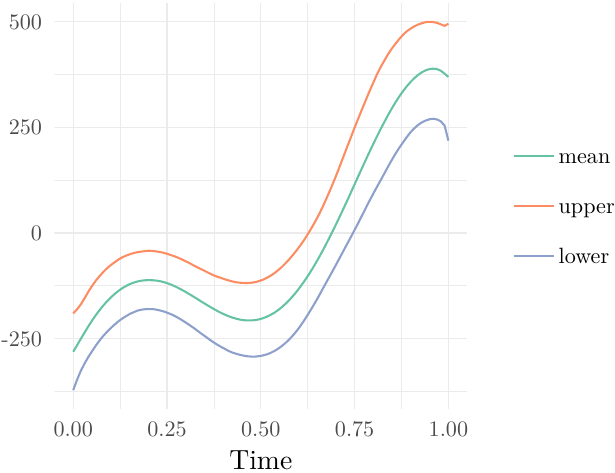} }

}

\caption{Tolerance bounds for the WRS axial contour weld dataset constructed using the methods in \cite{Rathnayake:2016} and \cite{Lewis:2017}.}\label{fig:weld_lewis_boot_amp}
\end{figure}

Figure \ref{fig:weld_boot_amp} presents the tolerance bounds for (a)
amplitude and (b) phase. The tolerance bounds were again calculated
using 500 bootstrap resamples with a sample size of 5 functions in each
iteration. We chose the two leading principal directions of variability as they
captured over 90\% of the overall variability. Both tolerance bounds have 99\%
coverage with a 95\% confidence level. Figure
\ref{fig:weld_boot_surf_amp} shows the surface plot of the tolerance
bounds for the phase and amplitude components. This is the same dataset
used in \cite{Lewis:2017}, and it is clear that our tolerance bounds
are more representative of the shape in the original data. Furthermore,
the separate bounds on phase and amplitude show the overall contribution
of each source of variability. This impact is lost in the tolerance
bounds generated by Lewis et al. \cite{Lewis:2017} as shown in Figure
\ref{fig:weld_lewis_boot_amp}(b).

Figure \ref{fig:weld_lewis_boot_amp}(a) presents the tolerance bounds
calculated using the approach of Rathnayake and Choudhary \cite{Rathnayake:2016}. The first peak and valley in these bounds are not as sharp as those computed using the proposed method and show in Figure \ref{fig:weld_boot_amp}(a). Figure
\ref{fig:weld_lewis_boot_amp}(b) presents the tolerance bounds
calculated using the approach of Lewis et al. \cite{Lewis:2017}. While the features of these bounds are sharper than those in Figure \ref{fig:weld_lewis_boot_amp}(a), they fail to capture the phase variability in the given data. It also appears that the main source of variability in these bounds is simply vertical translation, with very little variability in the shape.

\begin{table}[!t]
\centering
\begin{tabular}{ccccc}
  \hline
 & Tolerance Factor & No of PCs & Tolerance score mean (sd) & Sample Size \\
  \hline
Weld & 154.96 & 2 & 1.60 (0.80) & 5 \\
  ECG & 17.29 & 4 & 3.95 (2.68) & 80 \\
   \hline
\end{tabular}
\caption{Summary of results obtained using the combined fPCA-based tolerance region approach.}
\label{tab:pca_results}
\end{table}

The first row in Table \ref{tab:pca_results} presents the calculated
tolerance factor for this dataset for a tolerance region with 99\%
coverage and 95\% confidence. As before, we again used two principal directions of variability in the combined fPCA model to compute the tolerance factor. In this case, since the dataset contains only
five samples, the tolerance factor is quite large due to sampling
uncertainty. The mean of the tolerance scores computed for each of the
functions in this data is smaller than the tolerance factor. However,
with a larger sample size, we would expect the tolerance factor to be much smaller.

\hypertarget{pqrst-complexes-from-ecg-signals}{%
\subsection{PQRST Complexes from ECG
Biosignals}\label{pqrst-complexes-from-ecg-signals}}

\begin{figure}[!t]

{\centering \includegraphics[width=5.5in]{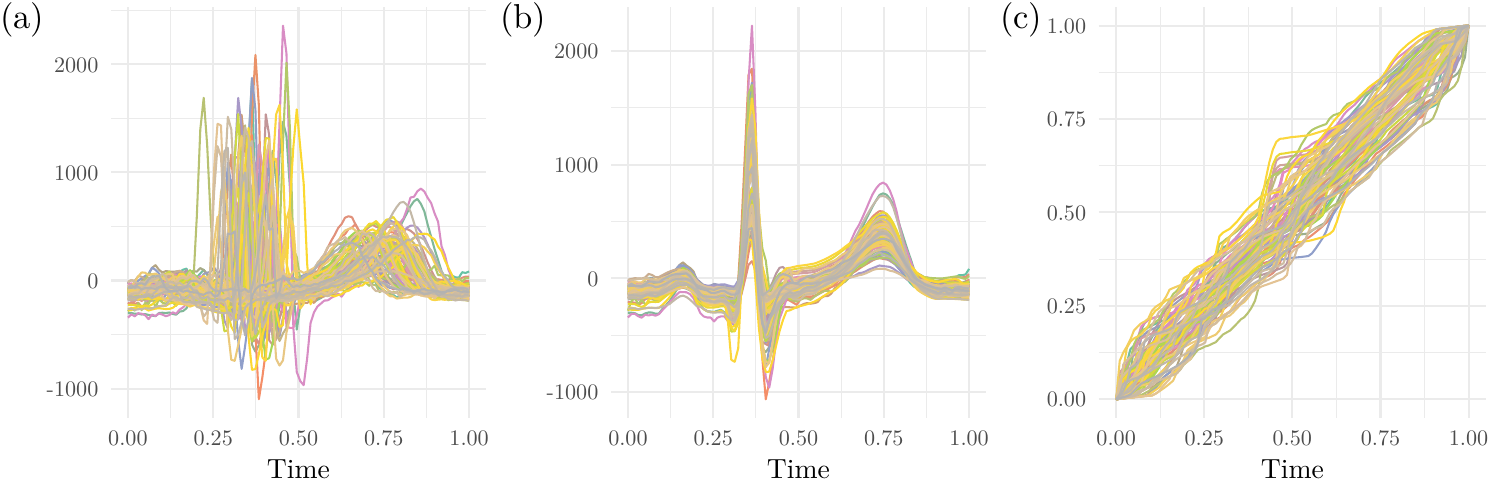}

}

\caption{Alignment of the PQRST ECG data. (a) Original functions. (b) Aligned functions (amplitude). (c) Warping functions (phase).}\label{fig:ECG_data}
\end{figure}

The electrocardiogram (ECG) data used in this work was obtained from the PTB Diagnostic ECG Database \cite{bousseljot:1995} on PhysioNet \cite{physiobank}. The ECG is a medical diagnostic tool that is routinely used to monitor the function of the heart, and is standard for diagnosing and monitoring various heart diseases and conditions, including myocardial infarction. The dataset considered in this work consists of 80 PQRST complexes segmented from a long ECG signal using the method presented in Kurtek et al. \cite{kurtek:2013}. Each complex corresponds to a single heartbeat where PQRST refer to the five peak and valley features (P=slight first peak, Q=sharp first valley, R=sharp second peak, S=sharp second valley and T=slight third peak). While each PQRST complex corresponding to a healthy heartbeat contains these five features, the magnitude of the peaks and valleys can be significantly different across subjects (i.e., amplitude variability); these magnitudes can also vary for different heartbeats within a single subject, albeit not to the same degree. Furthermore, the timing of these features can also be quite different (i.e., phase variability). Phase variability in this application corresponds to different timings and durations of heartbeats across individuals. This motivates the use of the proposed method to construct separate tolerance bounds for the amplitude and phase components based on this data. Figure \ref{fig:ECG_data}(a) displays the 80 segmented PQRST complex functions. It is clear that there is significant phase variability in this data as the five features are not well-aligned. The aligned functions (amplitude) and corresponding warping functions (phase) are shown in panels (b) and (c), respectively.

\begin{figure}[!t]

{\centering \subfloat[Amplitude\label{fig:ECG_boot_amp1}]{\includegraphics{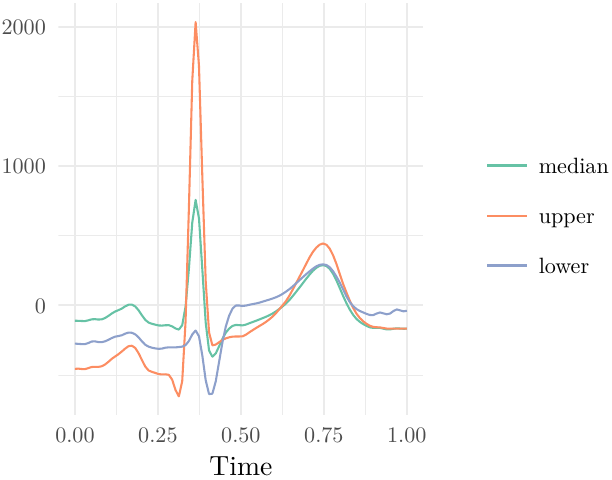} }\subfloat[Phase\label{fig:ECG_boot_amp2}]{\includegraphics{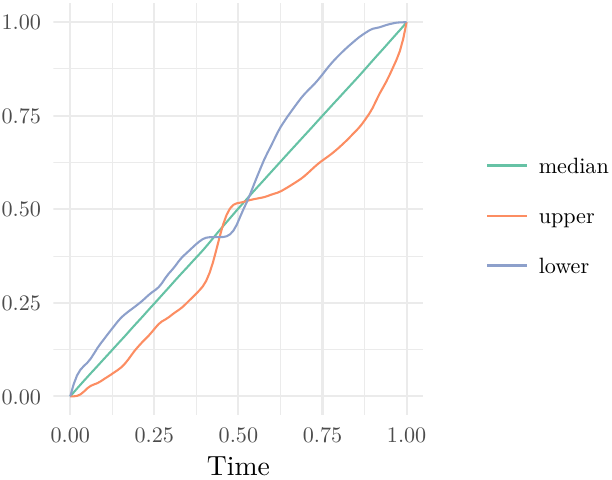} }

}

\caption{Bootstrapped geometric tolerance bounds for the PQRST ECG dataset.}\label{fig:ECG_boot_amp}
\end{figure}

\begin{figure}[!t]

{\centering \subfloat[Amplitude\label{fig:ECG_boot_surf_amp1}]{\includegraphics{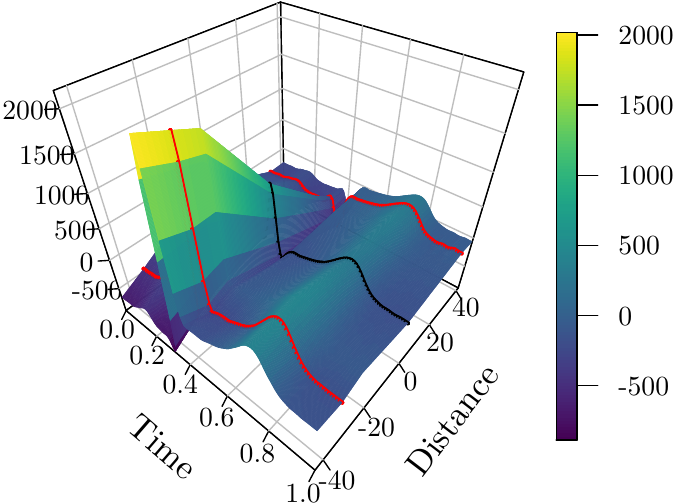} }\subfloat[Phase\label{fig:ECG_boot_surf_amp2}]{\includegraphics{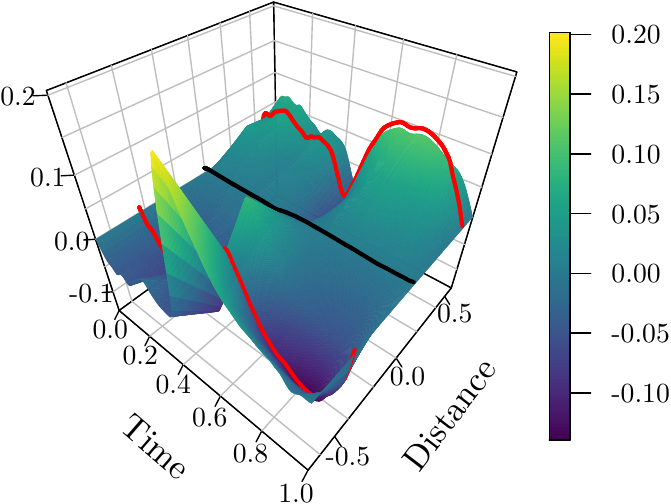} }

}

\caption{Surface plots of the bootstrapped geometric tolerance bounds for the PQRST ECG data.}\label{fig:ECG_boot_surf_amp}
\end{figure}

Figure \ref{fig:ECG_boot_amp} presents the bootstrapped tolerance bounds
for (a) amplitude and (b) phase. We again use 500 bootstrap resamples
with a sample size of 50. The number of principal directions of variability chosen from the combined fPCA model was
four; these directions captured over 90\% of the overall variability in the data. Both tolerance bounds
have 99\% coverage with a 95\% confidence level. Figure
\ref{fig:ECG_boot_surf_amp} shows the corresponding surface plots. The
amplitude tolerance bounds capture the relative sizes of the three peaks
and two valleys. This is well-demonstrated in the surface plot in Figure
\ref{fig:ECG_boot_surf_amp}(a). The phase tolerance bounds exhibit
the variability in the location of the three peaks.

\begin{figure}[!t]

{\centering \subfloat[Rathnayake and Choudhary \cite{Rathnayake:2016}\label{fig:ecg_lewis_boot_amp1}]{\includegraphics{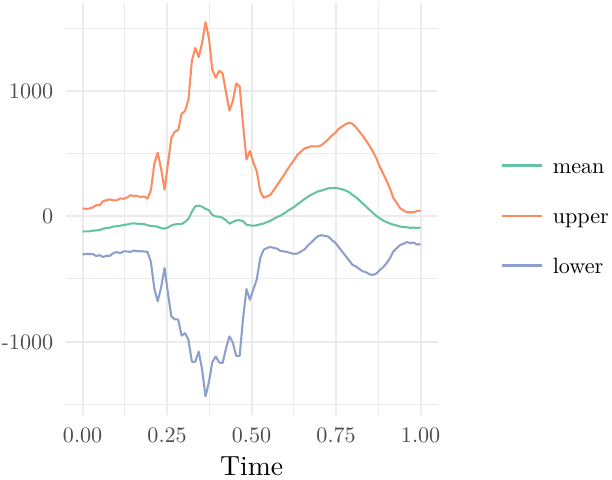} }\subfloat[Lewis et al. \cite{Lewis:2017}\label{fig:ecg_lewis_boot_amp2}]{\includegraphics{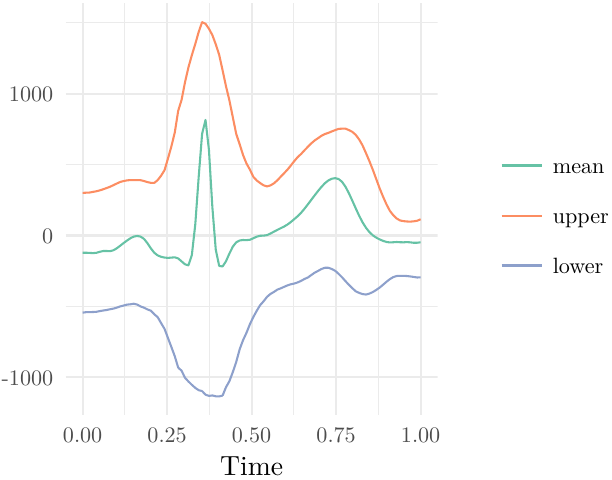} }

}

\caption{Tolerance bounds for the PQRST ECG dataset constructed using the methods in \cite{Rathnayake:2016} and \cite{Lewis:2017}.}\label{fig:ecg_lewis_boot_amp}
\end{figure}

Figure \ref{fig:ecg_lewis_boot_amp}(a) presents the tolerance bounds
calculated using the approach of Rathnayake and Choudhary \cite{Rathnayake:2016}. The bounds computed using this approach completely lose the structure present in the original data. This is precisely due to the fact that there is considerable phase variability in the given PQRST ECG signals. As before, it is difficult to determine the
relative contributions of amplitude and phase to the computed tolerance bounds,
and what the lower bound actually means in terms of the semantic features of the PQRST complex. Figure \ref{fig:ecg_lewis_boot_amp}(b) presents the tolerance bounds
calculated using the approach of Lewis et al. \cite{Lewis:2017}. Qualitatively, these bounds appear better than those in panel (a). However, comparing these bounds with those presented in Figure \ref{fig:ECG_boot_amp}, we still see major distortions of the PQRST features.

The last row in Table \ref{tab:pca_results} reports the calculated
tolerance factor for the PQRST ECG dataset for 99\% coverage with 95\%
confidence. We used four principal directions of variability in the combined amplitude-phase fPCA model to compute the tolerance factor. Again, the mean tolerance score of the functions in this dataset is smaller than the computed tolerance factor.

\hypertarget{conclusion}{%
\section{Discussion and Future Work}\label{conclusion}}

We presented two methods for computing tolerance bounds for elastic
functional data, i.e., functional data with random warping variability.
For both methods, we used a combined amplitude and phase functional
Principal Component Analysis model. The fPCA was used to define a
convenient generative model, which is easy to sample from. This enabled
the implementation of an efficient bootstrapping procedure to generate
geometrically-motivated tolerance bounds. The second approach used the
multivariate Gaussian fPCA model directly to define tolerance regions,
and to compute a corresponding cutoff value called the tolerance factor.
Therefore, one can easily test whether a function falls inside or
outside this tolerance region by computing a simple tolerance score on the fPCA coefficient space. We demonstrated the applicability of these two
approaches on a simple simulated example as well as two real data
examples wherein the observed functional data has clear amplitude and phase variation.

In this work, we have focused on accounting for phase and amplitude variability.
However, pointwise noise is always troublesome for the separation of phase and amplitude in functional data. Furthermore, the preprocessing required to create functions from raw data can also have an impact on the constructed tolerance
bounds. The proposed method relies on the assumption that the input functions are at least absolutely continuous. However, the numerical procedures used in this work have better behavior for smoother functional data. Thus, care needs to be taken in this aspect. In future work, we plan to quantify the robustness of the proposed procedure to various preprocessing steps.

Since the proposed methods rely on fPCA to construct tolerance bounds, certain
irregularities that fall outside of the space spanned by the leading fPCA basis functions
are ignored. In practice, one should select a sufficient number of fPCA basis
functions to capture all relevant directions of variability; if one
believes that small scale variability is important, then even basis
functions with small eigenvalues should be included in the analysis. On
the other hand, if the small scale irregularities are precisely the ones
that should be flagged, then leaving them out during the construction
of the tolerance bounds is beneficial. While we have outlined some approaches to selecting an appropriate number of fPCA basis functions to define the tolerance bounds, our future work will focus on the effects of this choice in real applied settings.

We have additionally identified several other directions for future work. First, we
will explore the influence of the weight \(C\) in the combined amplitude
and phase fPCA model on the resulting tolerance bounds and tolerance
factor. In particular, we want to assess the effects on simulated
confidence values. Second, our method relies on tangent space approximations for the phase component,
and recently Yu et al. \cite{yu:2017} showed that in some cases the method of
Principal Nested Spheres for dimension reduction provides more intuitive results. Third,
in many applications, the functional data of interest may be more
complex than the simple univariate functions considered in this work;
some examples include shapes of curves, surfaces, and images. These more
complicated data objects often exhibit additional sources of variability beyond amplitude and phase,
which must be taken into account when computing tolerance bounds.

\hypertarget{acknowledgment}{%
\section*{Acknowledgments}\label{acknowledgment}}
\addcontentsline{toc}{section}{Acknowledgment}

This paper describes objective technical results and analysis. Any
subjective views or opinions that might be expressed in the paper do not
necessarily represent the views of the U.S. Department of Energy or the
United States Government. This research was in part supported by the
National Technical Nuclear Forensics Center (NTNFC) of the U.S.
Department of Homeland Security (DHS). Sebastian Kurtek's work was partially supported by NSF grants DMS-1613054, CCF-1740761 and CCF-1839252, and by NIH grant R37 CA214955. The authors would like to thank Dr. Marc
Welliver at Sandia National Laboratories for his technical support during this work. They would also like to acknowledge the Associate Editor and Reviewer for providing constructive comments that have significantly improved the content of this manuscript.

\bibliographystyle{tfs}
\bibliography{JDTBib}

\end{document}